\documentclass[useAMS,usenatbib]{mn2e}
\usepackage{graphicx}
\usepackage{float}
\usepackage[draft]{hyperref}

\newcommand{\msun}{M_{\odot}}

\title[Stellar haloes reflect merger histories]{The masses and metallicities of stellar haloes  
reflect galactic merger histories}

\author[D'Souza \& Bell]{Richard D'Souza$^{1}$ $^{2}$, 
\thanks{Contact e-mail:\href{mailto:radsouza@umich.edu}{radsouza@umich.edu}}
						Eric F.\ Bell$^{1}$\\
$^{1}$University of Michigan, Department of Astronomy, 311 West Hall, 1085 South University Ave., Ann Arbor, MI 48109-1107\\
$^{2}$Vatican Observatory, Specola Vaticana, V-00120, Vatican City State
}

\begin{document}
\date{Accepted . Received ; in original form }
\pagerange{\pageref{firstpage}--\pageref{lastpage}} \pubyear{2017}

\maketitle

\label{firstpage}

\begin{abstract}
There is increasing observational and theoretical evidence for a correlation between 
the metallicity and the mass of the stellar halo for galaxies with Milky Way-like stellar 
masses.  Using the Illustris cosmological hydrodynamical simulations,  we find that this 
relationship arises because a single massive massive progenitor contributes the bulk 
of the mass to the accreted stellar component as well as sets its metallicity. Moreover, 
in  the Illustris simulations, this relationship extends over 3 orders of magnitude in 
accreted stellar mass for central galaxies. We show that for Milky Way-like mass galaxies, 
the scatter in accreted metallicity at a fixed accreted stellar mass encodes information 
about the stellar mass of the dominant accreted progenitor, while the radial density and 
metallicity gradients of the accreted stellar component encodes information about the time of  
accretion of the dominant progenitor.  We demonstrate that for Milky Way-like mass galaxies, 
the Illustris simulations predict that the metallicity and the stellar mass of the total accreted 
stellar component can be reconstructed from aperture measurements of the stellar halo along the 
minor axis of edge-on disk galaxies. These correlations highlight the potential for observational 
studies of stellar halos to quantify our understanding of the most dominant events in the growth 
history of galaxies. We explore the implications of our model for our understanding of the accretion 
histories of the Milky Way, M31 and NGC 5128. In particular, a relatively late and massive accretion 
is favoured for M31; additionally, we provide a first estimate of the accreted stellar mass for NGC 
5128.
\end{abstract}


\section{Introduction}
The stellar halo of a galaxy encodes information about its accretion history, and 
has been the {\it raison d'\^{e}tre} for the community to pursue the study of the 
faint surface brightness stellar haloes around galaxies. Today, we have amassed 
detailed resolved stellar maps of the halo of a handful of nearby galaxies 
\citep[for example][]{Mouhcine2010, Ibata2014, Crnojevic2016}. Yet, little 
progress has been made towards deciphering galactic accretion histories from 
these detailed observational maps of their stellar haloes. Nevertheless, as the 
number of observations of the stellar haloes of nearby galaxies have increased, 
we have begun to appreciate the diversity in the accretion histories of 
$\mathrm{L_*}$ galaxies \citep{Mouhcine2005b,Monachesi2016a,Merritt2016,Harmsen2017}. 

Hints of this diversity were first apparent in comparing the metallicity of the 
stellar halo of our own Milky Way (MW) with that of its neighbour, M31 \citep{Mould1986, 
Reitzel1998, Durrell2001, Ferguson2002, Durrell2004}. Despite comparable total galactic 
luminosities and stellar masses, the stellar halo of M31 has a mean metallicity which 
is nearly an order of magnitude higher than the stellar halo of the MW. Further evidence 
from neighbouring disk $\mathrm{L_*}$ galaxies indicated a range in metallicity in their 
stellar haloes \citep{Mouhcine2005b}.  With the exception of the MW, the metallicity of 
the stellar haloes of these galaxies correlates with the total galaxy luminosity 
\citep{Mouhcine2005b}. As a bright galaxy with an old metal-poor stellar halo, the MW 
appeared quite rare. 

Further evidence of this diversity came in recent measurements of the stellar mass of 
the halo of nearby disk $\mathrm{L_*}$ galaxies \citep{Trujillo2016,Merritt2016,Harmsen2017}. 
Indeed, the masses of most of the stellar haloes of nearby disk galaxies \citep{ 
Merritt2016,Harmsen2017} lie between the extremes charted out by the Milky Way \citep{Bell2008} 
and M31 \citep{Ibata2014}. 

With the help of early pioneering simulations, the diversity in the metallicity of the 
stellar haloes was interpreted in terms of differences in their assembly histories 
\citep{Renda2005,Font2006}. \cite{Renda2005} showed that galaxies with more extended 
merging histories possessed stellar haloes with younger and more metal rich stellar 
populations than the stellar haloes of galaxies with a more abbreviated assembly history
\citep[see also][]{Purcell2008}. Moreover, these simulations predicted a relationship 
between the metallicity and the mass of the stellar halo. Using the particle-tagging 
simulations of \cite{Bullock2005}, \cite{Font2006} also demonstrated a relationship
between the metallicity and stellar mass of their accreted haloes of their 11 disk-like 
galaxies, showing that the relationship mimicked the relationship between the 
metallicity-stellar mass relationship of dwarf galaxies \citep{Gallazzi2005, Woo2008} 
which were destroyed to build up the accreted stellar halo. Based on these models, 
\cite{Robertson2005} proposed that the diversity in the metallicities of the stellar 
haloes of nearby galaxies  reflected the diversity in the mass of the most massive 
progenitors accreted by these galaxies. Recently, \cite{Deason2016} using cosmological 
accretion-only models of a much larger number of MW-mass haloes confirmed these ideas 
and showed that the most massive progenitor, with $\mathrm{M_{*}} \sim 10^{8}-10^{10} \msun$, 
contributes the bulk of the accreted stellar material and drives the correlation between 
the metallicity and the mass of the stellar halo. 

Recently, \cite{Harmsen2017}, using the GHOSTS (Galaxy Halos, Outer disks, Substructure, Thick disks, 
and Star clusters) survey \citep{Radburn2011,Monachesi2016a}, demonstrated for the first time 
observationally a relationship between the metallicity of halo stars (measured at 30 kpc along 
the minor axis) and the mass of the stellar halo of 6 nearby MW-size edge-on disk galaxies (plus 
the MW and M31).  Given the predicted variations in the accretion histories of 
MW-like galaxies, the existence of such a relationship between the metallicity and 
the mass of the stellar halo of nearby galaxies is truly remarkable, and suggests that these 
Hubble Space Telescope (HST) observations along the minor axis of these nearby MW-peers are 
able to probe the characteristics of the most massive progenitor accreted by these galaxies. 
This opens the possibility of empirically exploring the relationship between the most massive 
merger/accretion partner and galactic properties \citep[e.g.,][]{Bell2017} --- a potentially 
crucial insight into how merging and accretion affect galaxies.

In light of this new observational evidence and additional constraints on the galaxy metallicity-
stellar mass relationship for dwarf galaxies \citep{Kirby2013}, it is appropriate to revisit 
the modelling of the relationship between the metallicity and the stellar halo mass for a 
range of galaxy types and sizes, in order to understand how informative this relationship 
is about the accretion history of the galaxy, especially in terms of the characteristics of 
the most massive accreted progenitors. The availability of large cosmological hydrodynamical 
simulations \citep[e.g. Illustris and EAGLE;][]{Vogelsberger2014a,Schaye2015} allows us to study 
the correlation between the metallicity and mass of the stellar halo of galaxies for a broader 
range in DM halo masses and accretion  histories, while the spatial resolution of these 
simulations allows us to connect these quantities more directly to observable ones.


In this study, we use the GHOSTS HST observations as a template for possible observable
quantities. The GHOSTS survey targeted nearby edge-on disk galaxies with pencil beam 
observations, where stellar halos were probed using pointings along the minor axis (or axes) 
with galactocentric distances  $5<d_{\rm minor\,axis}<80$\,kpc, and more typically 
$10<d_{\rm minor\,axis}<80$\,kpc. Additionally, a power-law density profile and a 
metallicity gradient of the stellar halo can also be quantified at this distance. 
Comparing these observations with accretion-only models, \cite{Harmsen2017} concluded
that the halo stars probed at these distances were predominately of accreted origin.

However, hydrodynamical simulations also predict the presence of {\it in situ} stars 
(born within the galaxy and not accreted) at large galacto-centric distances 
\citep{Zolotov2009, Font2011, Pillepich2015}, commonly called as the `{\it in situ}' 
stellar halo. Although quantitative predictions of the amount of contribution 
of {\it in situ} stars to the stellar halo are highly dependent on numerical (e.g., 
initial conditions, accretion history) and modelling (e.g., prescriptions for 
star formation and the treatment of the multi-phase interstellar medium) techniques 
employed in the simulations \citep{Zolotov2009,Cooper2015}, there is a consensus 
within the community that the stellar halo beyond 30 kpc is dominated by the accreted 
stellar component \citep{Cooper2015}. The Illustris  simulations  predict a relatively 
large fraction of {\it in situ} stars at large galacto-centric distances beyond 30 kpc
\citep{Pillepich2014}, even along the minor axis. These predictions appear incompatible  with present  GHOSTS observations \citep[see also section \protect\ref{subsec:comparison}]{Harmsen2017}. 

In this work, we instead focus on the accreted stellar component of hydrodynamical  simulations. These accreted stars are tidally torn from accreted dwarf galaxies.
The ingredients that are vital to building up the accreted stellar component of a galaxy are relatively realistic and simple by design. Consequently, any model that has {\it i) } a realistic mass function of galaxies, {\it ii)} reasonable merger histories, and {\it iii)} accurately modelled potential wells will produce realistic accreted halos. These requirements are satisfied in present day hydrodynamical simulations. Hence, the accreted stellar component is fairly robustly modelled in the present set of hydrodynamical simulations. On the other, because the \emph{in situ} stellar component is highly dependent on poorly constrained recipes for star formation and stellar feedback in the outer lower density regions of a galaxy, they are extremely sensitive to a number of modelling dependencies and vary considerably from simulation-to-simulation.

In this work, we also focus on the total accreted stellar component, as 
opposed to only the accreted stellar material beyond a certain galacto-centric radius. We adopt 
this approach in order to examine those physical quantities that are most informative about 
the accretion history of the galaxy. Yet, given that much of the accreted material is 
expected to have settled into the centre of the galaxy where it is observationally 
difficult to access, even for minor axis resolved stellar populations studies, we 
will explore also how minor axis properties at larger radii relate to the total 
accreted quantities. Furthermore, for MW-like galaxies, we assume that the 
stellar halo probed at 30 kpc along the minor axis (GHOSTS HST observations) is predominately 
of accreted origin. 

In Section \ref{sec:Ill}, we review the suitability of using the accreted stellar 
component of the Illustris galaxies to study the stellar haloes of MW-mass galaxies.
In Section \ref{sec:rel}, we demonstrate that the bulk of the accreted stellar component
is built up by the most massive accreted progenitor, and that it drives the accreted  
metallicity-stellar mass relationship. In Section \ref{sec:scatter}, we show how the 
scatter in the accreted metallicity relationship is informative about the fraction of 
accreted stellar mass contributed by the dominant progenitor. In Section \ref{sec:MW}, we 
explore how the physical properties (e.g, density or metallicity gradients) of the 
accreted stellar component of MW-like mass galaxies encodes information about the 
time of accretion of the dominant progenitor. In Section \ref{sec:minor}, we demonstrate
how it is possible to reconstruct the mass and metallicity of the total accreted 
stellar component from observable ``aperture" measurements. Finally, in Section 
\ref{sec:discuss}, we discuss our results and their implications for the MW, M31 and Cen A.

\section{Illustris Simulations}
\label{sec:Ill}
We use the public release of the Illustris suite of simulations \citep[which are described 
in detail in][]{Vogelsberger2014a,Genel2014,Vogelsberger2014b}. These are large cosmological 
hydro-dynamical simulations (with a periodic box of 106.5 Mpc on a side) run with the 
\texttt{AREPO} code \citep{Springel2010} and  include key physical processes  that  are 
believed to be relevant for galaxy formation \citep{Vogelsberger2013} including gas 
cooling with radiative self-shielding corrections,  star formation, energetic feedback 
from growing SMBHs and exploding supernovae, stellar evolution with associated chemical 
enrichment and stellar mass loss, and radiation  proximity effects for AGNs. The Illustris 
simulations produce a morphologically  diverse galaxy population, reproducing median 
morphology trends with stellar mass, SFR and compactness \citep{Snyder2015}. The details 
of the public release are found in \cite{Nelson2015}.

We use Illustris-1, the highest resolution run, which follows the  evolution of dark matter (DM), 
cosmic gas, stars and super massive black holes from a  starting redshift of  $z=127$  
to $z=0$ with cosmological parameters consistent with WMAP-9 results. This run has a DM
mass  resolution of  $\mathrm{m_{DM}}=6.26 \times 10^{6} \msun$ and a resolution of 
$\mathrm{m_{baryon}}= 1.26  \times 10^{6} \msun$ for the baryonic component. The 
gravitational softening lengths are 1.4  and 0.7 kpc for the DM and baryonic  
particles respectively at $z=0$. Haloes, subhaloes and their basic properties have been 
identified using the Friends-of-Friends (FoF) and SUBFIND algorithms 
\citep{Davis1985,Springel2001,Dolag2009}. We use the merger trees as identified by 
the {\sc SubLink} code  \citep{Rodriguez-Gomez2015}. 

In this work, we use a Hubble constant of $\mathrm{H_{0}=72\,km\,s^{-1}\,Mpc^{-1}}$. 
For the virial mass of a particular galaxy, we use the total mass of the group enclosed 
in a sphere whose density is 200 times the critical density of the Universe (contained 
in $\mathrm{Group\_M\_Crit200}$). For the stellar mass of a particular galaxy, we use 
the stellar mass of all the stellar particles bound to this subhalo (contained in $\mathrm{SubhaloMassType}$). 

We use the metallicity values (all elements above He) of the stellar particles as 
found in the snapshots  and the halo catalogues. These metallicity values are inherited 
from the gas cell which get converted into stars at the time of their birth. The 
metallicity values found in the Illustris simulations are a product of their choice of 
stellar evolution and metal  enrichment. Furthermore, no metallicity  floor has been  
imposed to the output data, so that the metallicities of a small fraction of gas  
and star elements adopt minuscule, unrealistic values. We  impose a lower limit to 
the stellar metallicities in this work ([M/H] $\sim$ $-$5.4). The median fraction of 
stellar particles affected is 0.1\%, with an upper limit of 0.2\% for smaller 
galaxies. We find that these stellar particles with unrealistic metallicities
have a minimal effect on the median accreted stellar metallicity of galaxies. 
We choose to retain these stellar particles with a lower limit in stellar 
metallicities, so as not to bias our estimates of the accreted stellar mass.

\begin{figure}
\centering
\includegraphics[width=80mm]{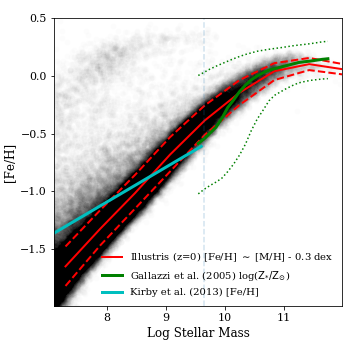}
\caption{The mass-weighted mean metallicity of Illustris central galaxies as a 
function of the galaxy  stellar mass. The red solid line and dashed lines indicate 
the median and the 16/84 percentile of the distribution. The metallicity 
of the Illustris galaxies is lowered by 0.3\,dex to convert it to [Fe/H].
Observational constraints: The solid green line indicates the metallicity of SDSS galaxies 
as estimated by \protect\cite{Gallazzi2005}. The dotted green lines indicate 
the 16th and 84th percentile of the distribution. The solid cyan line are 
observational constraints of dwarf galaxies from \protect\cite{Kirby2013}. 
The Illustris mass-metallicity relationship is reasonably consistent and reproduces
the broad contours of the observational data.}
\label{fig1}
\end{figure}

The Illustris metallicity model is dependent upon the sub-grid ``physics'' models (e.g., 
stellar evolution, chemical enrichment of SN Ia/II and AGB stars, as well as AGN 
feedback and stellar winds) implemented in the simulation. Given these uncertainties, 
it is imperative that we compare the metallicity of the Illustris  galaxies 
as a  function of  their stellar mass (both overall normalisation and shape of the
relationship) with the relevant observational constraints available in the literature. 
We do so in Figure \ref{fig1}: the stellar  metallicity of SDSS galaxies 
\citep{Gallazzi2005} and of dwarf satellites of the Milky Way and M31 \citep{Kirby2013}. 
While \citealt{Gallazzi2005} measure metallicities using broad spectral features, 
dominated  by Mg and Fe in the integrated light of galaxies,  \citealt{Kirby2013} 
measure [Fe/H] from iron absorption lines in individual stars. For the purpose of 
this paper, we treat them equivalently as tracing predominately [Fe/H]. On the 
other hand, the Illustris model traces the total stellar metallicity [M/H] (all 
elements above He). In this work, we lower the metallicities of the Illustris 
galaxies by 0.3 dex to convert it to a [Fe/H] system and adjust for any 
limitations in the Illustris chemical evolution model. Doing so provides a 
a good match between the overall normalisation of the metallicity of the Illustris 
galaxies and the observational constraints.

The mean metallicity of the Illustris galaxies increases with stellar mass in reasonable 
accord with observations, although the exact shape of the Illustris stellar mass-metallicity mass relationship 
appears to differ somewhat from the observations.  In general, the mass-metallicity relationship of 
SDSS $\mathrm{L_{*}}$ galaxies is steeper than that found in the Illustris simulations, 
while the mass-metallicity relationship of dwarf galaxies is shallower than found in 
the simulations. The Illustris predictions are a direct consequence of the choice of the Illustris 
stellar evolution model as well as the recipes of galactic winds and AGN feedback employed 
in the simulation.  The slope of the mass-metallicity relationship of galaxies will 
have a direct impact on the accreted stellar metallicities derived later. 

\begin{figure}
\centering
\includegraphics[width=80mm]{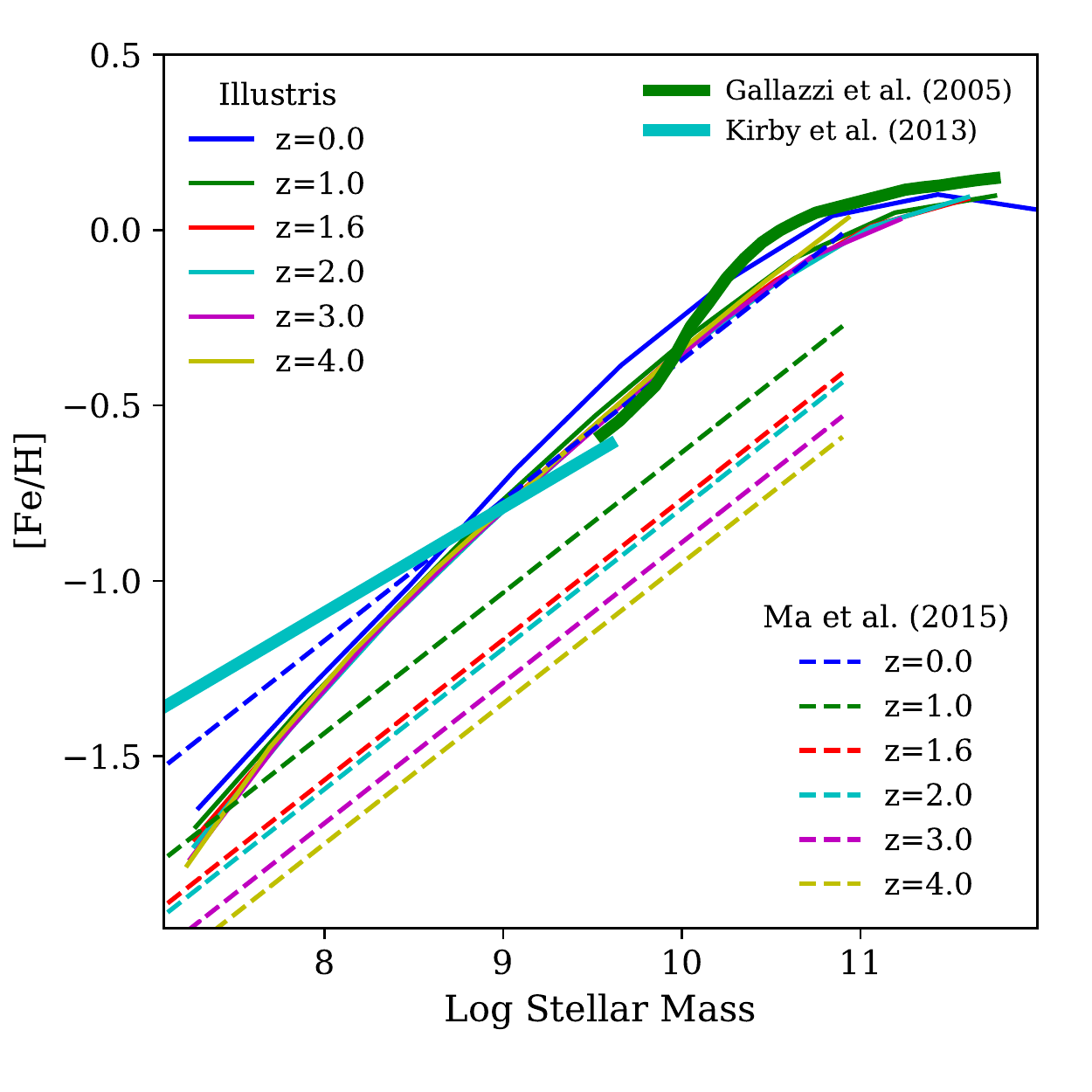}
\caption{The lack of evolution in the metallicity--stellar mass relationship of Illustris 
galaxies from  redshift z=4 to the present day universe (solid thin lines). We 
also compare this with evolution of galaxy metallicity--stellar mass relationship 
in the simulations of Ma et al. 2015 (dashed thin lines). For comparison, we also 
show the observational results of the metallicity-stellar mass relationship of 
local dwarfs of  Kirby et al. (2013) and of SDSS galaxies from Gallazzi et al. (2005).
}
\label{fig2}
\end{figure}

The Illustris simulations shows weak evolution of the stellar metallicity--stellar mass 
relationship with redshift with redshift  (See Figure \ref{fig2}). This behaviour is not 
inevitable: the simulations  of \cite{Ma2016} exhibit a strong time evolution in the 
stellar metallicity--stellar mass relationship. At a fixed stellar mass, they predict a  
strong evolution in stellar metallicity of nearly 0.5 dex  from $z\sim 3$ to  
$z \sim 0$.\footnote{The simulations of \cite{Ma2016} attribute the evolution in 
stellar metallicity with time at a fixed stellar mass to their different rates of 
star formation over cosmic time. In their simulations, the fraction of stellar 
material  to the total number of baryons (including the halo gas) increases with time.  
In other words, the suppression of star formation at high redshift leads to a decrease in the 
amount of  metals produced in galaxies. In contrast, in other simulations 
galaxies tend to form a large fraction of their stars at higher redshift.}

There are few observational constraints on the evolution of the stellar 
metallicity--stellar mass relationship with redshift. Although gas-phase 
metallicities have been measured to redshift $z\sim 2$ using gas-phase oxygen 
and nitrogen abundances of the interstellar medium, future estimates of the 
stellar metallicities of high redshift galaxies will provide crucial data for 
our present understanding of chemical evolution models which have not been 
reliably assessed outside the local Universe. Determining stellar metallicities 
for galaxies at high redshift is a difficult task and requires deep spectroscopy 
in order to trace the stellar continuum and the strength of key absorption features 
chiefly sensitive to age and metallicity. A few studies have attempted to push to 
higher redshift ($z \sim 2$) by using rest-frame UV absorption features 
\citep{Rix2004,Halliday2008,Sommariva2012}. \cite{Gallazzi2014} using deep multi-object 
optical spectroscopy found very little evolution in the stellar metallicity between 
$z\sim0.7$ and $z\sim0.1$ (a difference of 0.13 dex). Based on the limited observational 
evidence available so far, we conclude that the Illustris chemical evolution model is 
more consistent with the observations than the simulations of \cite{Ma2016}.

Given the overall reasonable agreement between Illustris and the observed metallicities of 
galaxies, and cognizant of its uncertainties, we proceed to analyze the properties of 
Illustris central galaxies. We select 4644 central galaxies at $z=0$ with DM 
subhalo masses $\log_{10} \mathrm{M_{halo}/h} \ge 11.5$. These central galaxies are those 
that have the  most massive {\sc SubFind} subhaloes within a FOF group. We ensure that each 
galaxy is resolved by more than 3500 stellar particles per halo. These galaxies span
a range in accretion histories and galaxy morphologies. 

In this work, we identify a stellar particle as \emph{in situ} if the galaxy in which 
it formed lies along the ‘main progenitor branch’ of the galaxy \citep[{\sc SubLink} 
merger trees][]{Rodriguez-Gomez2015} in which it is currently found. The other stellar 
particles found in the subhalo of the main galaxy are tagged as ``accreted''. We note that 
our definition of ``accreted'' stars includes those stellar particles formed from gas 
stripped from satellite galaxies and are eventually bound to the main subhalo of the host 
galaxy. Our definition of \emph{in situ} versus accreted stars is consistent with 
\cite{Rodriguez-Gomez2016}. Our classification of stars formed from gas stripped from 
satellite galaxies is closer to \cite{Pillepich2015} than \cite{Cooper2015}. It is worth
noting that in simulations using the \texttt{AREPO} code like Illustris, it is 
difficult to trace the origin of the gas particle responsible for the creation of a 
stellar particle. Hence, it is difficult to distinguish stars formed from `stripped gas' 
(from satellites) as opposed to stars formed from `smoothly accreted gas'. Indeed stars 
formed from gas stripped from satellites after infall into the main galaxy often share 
some of the properties of stars born before infall. The definitions we adopt in this 
paper, especially with regards to stars formed from gas stripped from satellites, are 
consistent with our overall aim of studying the accretion history of the galaxy.

We define the accreted stellar mass of a galaxy as the mass of stars born external to the 
main progenitor branch of the galaxy ($\mathrm{M_{acc}}$). For a few star-forming 
galaxies, the subhalo finder algorithm {\sc SubFind} sometimes identifies star forming regions 
as separate subhaloes. We consider any such groupings of stellar particles identified 
by the {\sc SubFind} algorithm as separate subhaloes without any associated DM as part 
of the in-situ stellar component of the galaxy.

For each galaxy, we identify the  `dominant' progenitor which contributes the 
largest amount of accreted stellar material to the galaxy. We identify the mass of 
the dominant progenitor ($\mathrm{M_{dom}}$) as the maximum mass of the progenitor satellite 
before it becomes accreted onto the main progenitor branch of the galaxy 
(see {\sc SubLink} merger trees above). We also quantify the fraction of 
accreted stellar material which was contributed by the most dominant 
progenitor ($\mathrm{f_{Dom}}$). We note that due to stripping due to
tidal effects, in some cases, $\mathrm{M_{Dom}}$ may be larger than  
$\mathrm{f_{Dom}} \times \mathrm{M_{Acc}}$. Using the {\sc SubLink} merger trees,  we can also
identify the time of accretion of the progenitor by the main galaxy. This
corresponds to the time the {\sc SubFind} algorithm can no longer identify 
the stellar particles associated with a particular progenitor as a separate 
entity but bound to the main galaxy. Note that in this work, $\mathrm{t}$ refers to
cosmic time, or the age of the Universe. Hence, $\mathrm{t\,=\,0}$ corresponds 
to the start of the simulation, or the Big Bang.

We characterise the first moment of the accreted stellar metallicity distribution 
function (MDF) in terms of the mass-weighted mean and the median metallicity. 
We find that the mass-weighted mean metallicity of the accreted stellar 
component is at the most 0.15 dex higher than the median metallicity. 
For a few isolated cases ($<10$), very recent and large accretion events 
(`giant streams') bias the mass-weighted metallicity to much higher values. 
For the purposes of this paper,  we only consider the median metallicity 
of the accreted stellar component. We do not identify and remove recent accretion 
events that in observations may appear as `Sagittarius'-like streams or objects. 
Not only would removal of such accretion events be explicitly resolution-dependent, 
but also GHOSTS-like observational studies often cannot recognize and do not 
remove such substructure. 

To compare our results with the  GHOSTS observations of neighbouring 
edge-on spiral galaxies \citep{Harmsen2017}, we consider the accreted 
stellar particles in a narrow wedge along the minor axis. For minor axis 
properties, we orient each galaxy along its principal axis and estimate 
the metallicity of the accreted stellar component in two  projected wedges of 30 
degrees width along the minor axis between a galacto-centric distance of 
25 and 45 kpc. To increase the number of particles and smooth out sudden 
variations due to the presence of substructure, we include particles on both 
axes located within the diametrically opposed wedge. 

We also characterise the (observationally-inaccessible) size of the accreted stellar 
component in terms of 3D radial distance which contains 50\% of its stellar mass 
($\mathrm{R_{50\,Acc}}$). We estimate the projected radial density profile  
of the accreted stellar component along the minor axis, and estimate its power-law  
slope ($\Gamma_{\mathrm{Min\,Acc}}$). Finally, we also estimate the projected 
accreted stellar metallicity gradient  along  the minor axis.

\subsection{Comparison of the Illustris stellar halos with observations}
\label{subsec:comparison}
\begin{figure}
\centering
\includegraphics[width=0.5 \textwidth]{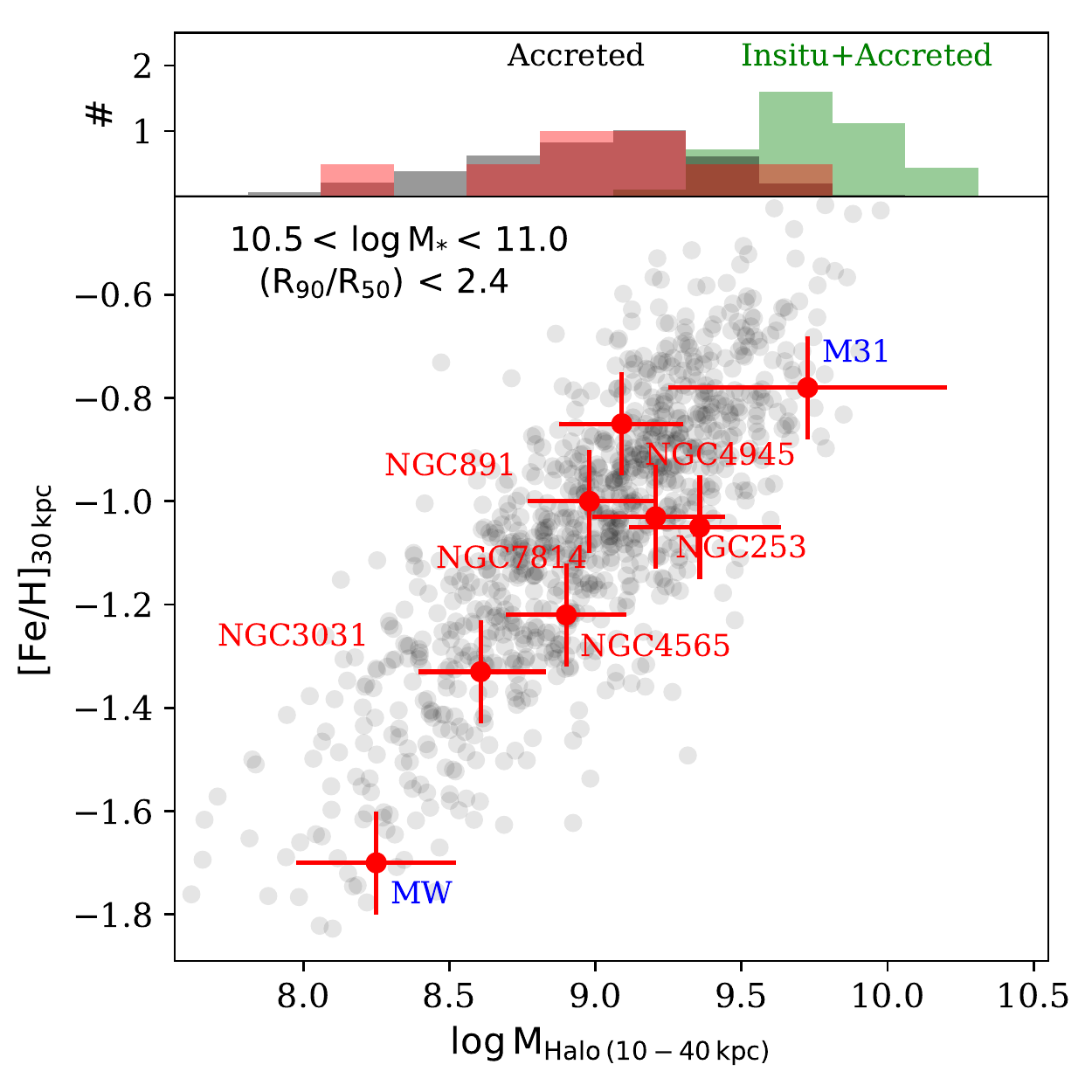}
\caption{{\bf Bottom Panel}: Accreted metallicity of the 6 GHOSTS galaxies 
(Harmsen et al. 2016) in addition to the Milky Way and M31 (plotted as 
a function of their `aperture' stellar halo mass measured between 10 
and 40 kpc).  Measurements of Milky Way and M31 are also taken from 
Harmsen et al. who used data from Bland-Hawthorn \& Gerhard 2016, 
Xue et al. 2015, Ibata et al. 2014, Gilbert et al 2014. The grey points 
indicated the accreted stellar metallicity estimated  in a wedge along 
the minor axis for Illustris GHOSTS-like galaxies ($10.5 <\, \log\, 
\mathrm{M_{*}} <\, 11.0$, $\mathrm{R_{90}/R_{50} \, < \,2.4}$) as a 
function of the accreted stellar mass measured in an aperture
between 10 and 40 kpc. {\bf Top Panel}: We compare the distribution 
function of the aperture accreted stellar masses of Illustris 
GHOSTS-like  galaxies (grey) with the aperture stellar halo masses 
of the  GHOSTS data + MW + M31 (red) measured between 10 and 40 kpc, 
finding that they are consistent with the accreted stellar component only.
In contrast, the distribution function of the aperture stellar halo 
masses (\emph{in situ} + accreted)  of Illustris GHOSTS-like galaxies 
is not consistent with the observational data.}
\label{fig18b}
\end{figure}

We now turn our attention to comparing how well the stellar haloes of the 
Illustris galaxies matches the observational data. We first examine the 
bulk properties of the stellar haloes of the GHOSTS galaxies and compare 
them with the accreted stellar component of the Illustris galaxies. In 
Fig. \ref{fig18b}, we show the estimated [Fe/H] at 30 kpc along the minor 
axis of  the 6 GHOSTS  galaxies \citep{Monachesi2016a}, along with the 
Milky Way and M31, as a function of their estimated stellar halo masses 
measured in an elliptical aperture measured between 10 and 40 \,kpc along the  
semi-minor axis \citep[as in Figure 12(b) of ][]{Harmsen2017}.

We compare these observations with the predictions of the accreted stellar 
component from the Illustris simulations. We select GHOSTS-like galaxies from 
the Illustris simulation by choosing low-concentration galaxies in a narrow 
stellar  mass range ($10.5 <\, \log\, \mathrm{M_{*}} <\, 11.0$, 
$\mathrm{R_{90}/R_{50}  \, < \,2.4}$). We estimate the median accreted stellar 
metallicity [Fe/H] in a narrow wedge along the minor axis for these GHOSTS-like 
Illustris galaxies (shaded points). We also estimate an aperture accreted stellar
mass between 15 and 50 kpc along the semi-minor axis, corresponding approximately 
to an elliptical aperture measured between 10 and 40 \,kpc along the minor axis.

Recall that the estimated [Fe/H] scaling that we have adopted for Illustris ([M/H]$-0.3$\,dex) 
gives a good match to the {\it galaxy} metallicity--stellar mass relationship 
(Fig.\ \ref{fig1}). With such an adjustment, the agreement between the Illustris 
predictions of the outer accreted stellar component and the observational constraints 
is quite remarkable. Furthermore, the distribution of the predicted aperture 
accreted  stellar mass (top panel of Figure \ref{fig18b}) matches the GHOSTS 
data well. In contrast, the distribution function of the aperture stellar halo 
masses (\emph{in situ} +  accreted; in green) of Illustris GHOSTS-like galaxies 
dramatically over-predicts the amount of stellar mass at large radii in the GHOSTS 
sample of galaxies.

\begin{figure}
\centering
\includegraphics[width=0.5 \textwidth]{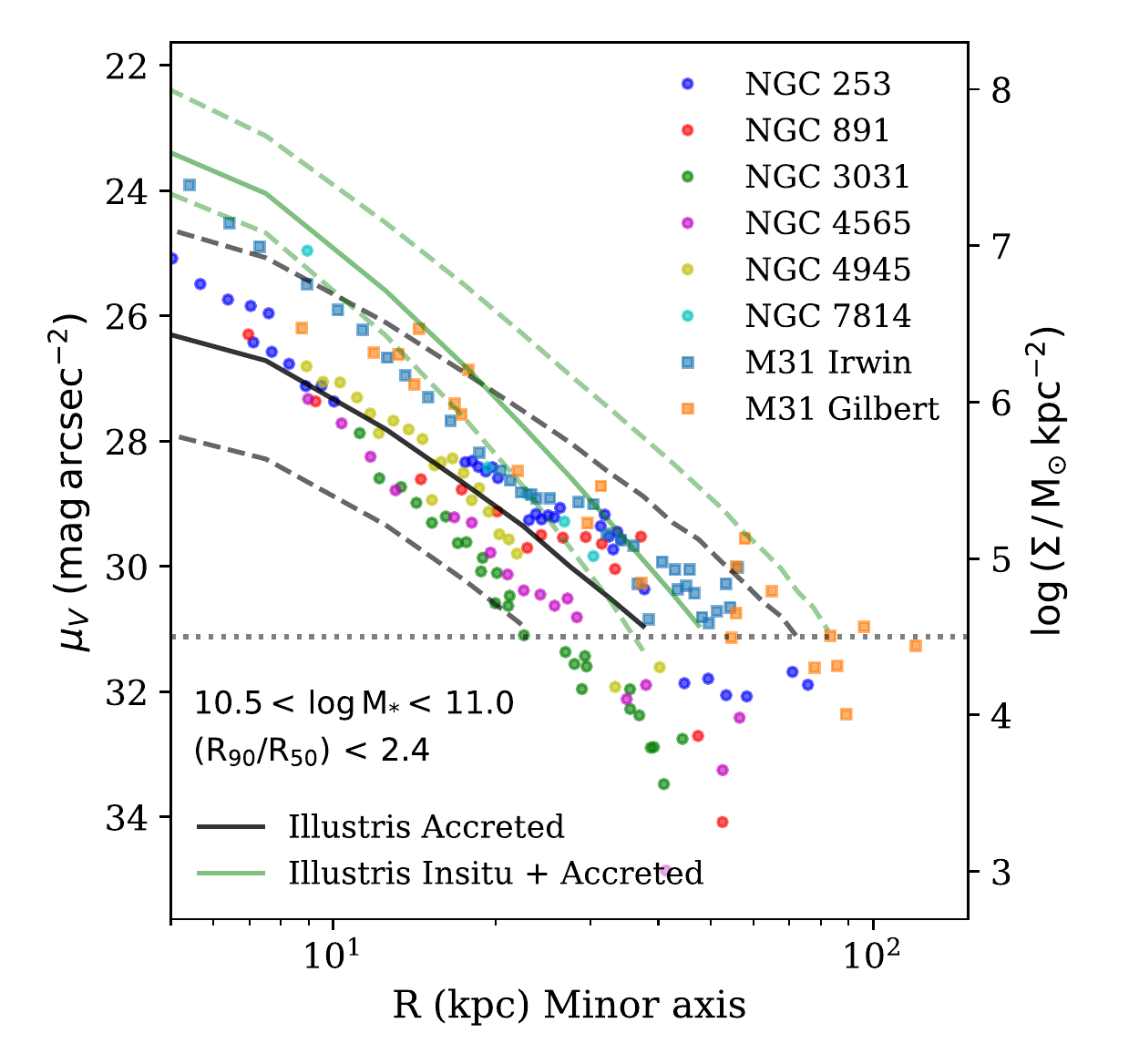}
\caption{Surface brightness profiles along the minor axis of the 6 
GHOSTS galaxies in the V-band from Harmsen et al. 2017. We also add the 
surface brightness profiles of M31 along the minor axis in the I-band from 
Irwin et al. 2005 and Gilbert et al. 2012. Assuming a M/L ratio of 2.5 in 
the V-band, we also plot a surface mass density profile of the total (green) 
and the accreted stellar component (black) along the minor axis for Illustris 
GHOSTS-like galaxies  ($10.5 <\, \log\, \mathrm{M_{*}} <\, 11.0$, 
$\mathrm{R_{90}/R_{50} \, < \,2.4}$). The dashed lines indicate the 16 and 84 
percentile of the radial distributions. Owing to the resolution of Illustris,
the surface mass density profiles do not extend lower than 
$\log_{10}\,\Sigma_{*}\,\sim\,4.5$  (indicated by the horizontal dotted line).
The surface mass density profiles of galaxies along the minor axis 
agree well with only the accreted stellar component beyond a radius of 15 kpc. 
Inclusion of the \emph{in situ} stellar component over-predicts the surface mass 
density profiles.}
\label{fig_GHOST_comp}
\end{figure}

Having demonstrated overall agreement with the bulk properties of the stellar
halo of the GHOSTS galaxies and the accreted stellar component of the Illustris
simulations, we proceed to a more stringent comparison of the surface mass density
profile of the GHOSTS galaxies along the minor axis, with the profiles predicted
from the Illustris simulations. In Figure \ref{fig_GHOST_comp}, we compare 
the estimated surface mass density profiles of the GHOSTS galaxies with the
profile of Illustris GHOSTS-like galaxies (selected above). We assume a constant
M/L ratio in the V-band of 2.5. We also include the observational constraints of 
the profiles of M31 along the minor axis \citep{Irwin2005,Gilbert2012} in our 
comparison. We find that the median surface density profile (along with the spread) 
of the accreted stellar component of Illustris GHOSTS-like galaxies matches the 
observational constraints quite well, especially between 10 and 50 kpc along 
the minor axis. On the other hand, the median surface density profile of the 
total (\emph{in situ} + accreted) is comparatively larger than the observational
constraints. Note that owing to the mass resolution of the Illustris simulations,
the surface mass density profiles do not extend lower than 
$\log_{10}\,\Sigma_{*}\,\sim\,4.5\,\mathrm{M_{\odot}/kpc^{2}}$.  This implies that the Illustris simulation over-predicts the  \emph{in situ} component along the minor axis. 
It also validates the claim of \cite{Harmsen2017} that the minor axis profiles of 
GHOSTS galaxies around 30 kpc are predominately of accreted origin.

The agreement between the Illustris accreted halo and observed halos at large radius 
is encouraging and supports the idea that exploring the accreted stellar component 
in Illustris could help build intuition about how to interpret the observed stellar halos. 
Accordingly, we now turn to a more detailed and systematic study of the Illustris 
accreted metallicity-stellar mass relationship.

\section{Accreted Metallicity-Stellar Mass Relationship}
\label{sec:rel}
\begin{figure}
\centering
\includegraphics[width=0.5 \textwidth]{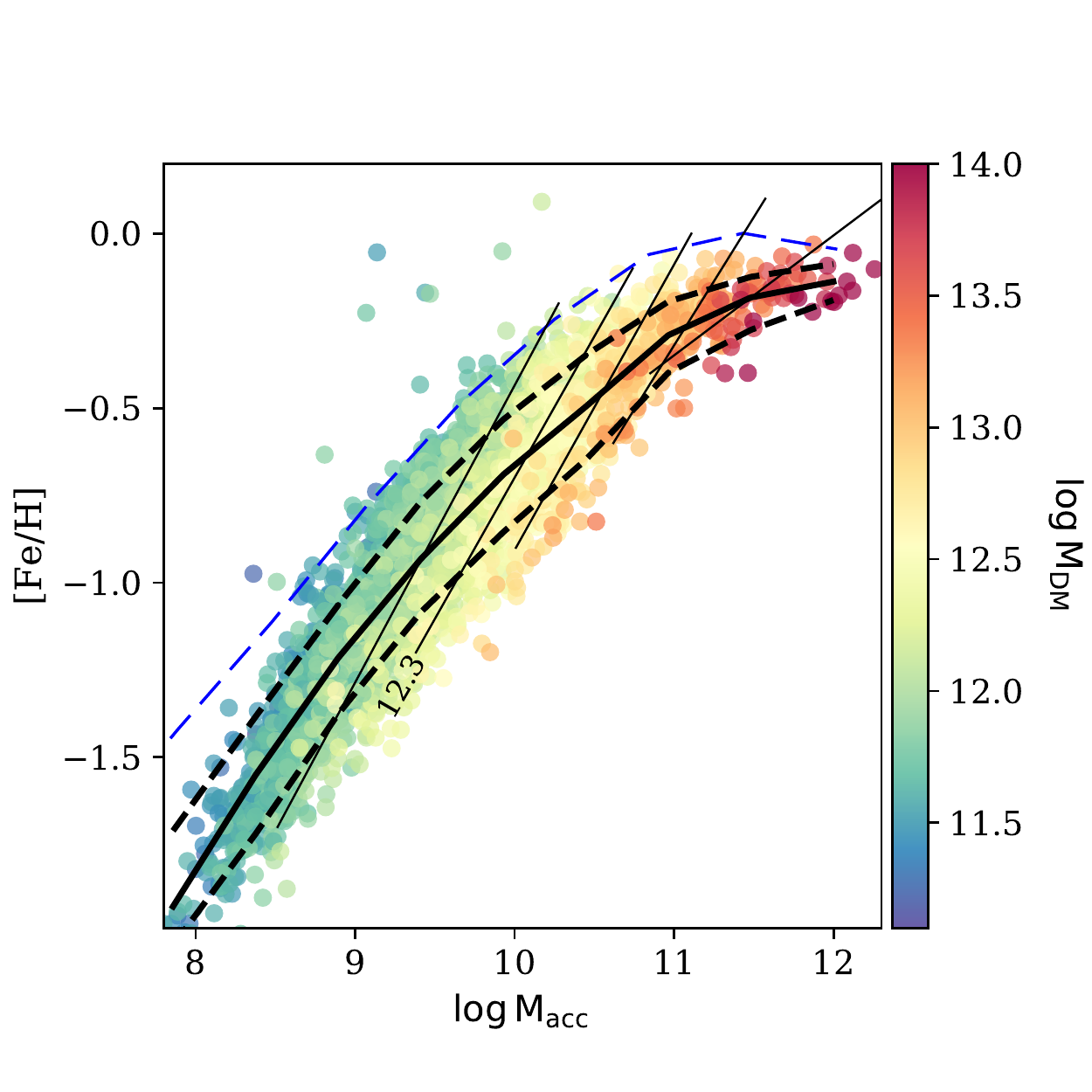}
\caption{ The Illustris accreted metallicity-stellar mass relationship: 
the metallicity of the accreted component is plotted as a function of 
the accreted stellar mass of the Illustris galaxies. The black solid 
and dashed lines indicate the median and the 16/84th 
percentile of the distribution.  The  galaxies are coloured by their 
halo mass. The thin black lines show tracks of similar
halo mass (at $\log_{10} \mathrm{M_{DM}}$ =  11.9,12.3,12.7,13.1 \& 13.5; 
discussed more in \S \ref{sec:MW}).
These contour-like solid lines are fitted to narrow ranges in 
$\log_{10} \mathrm{M_{DM}}$. We overplot the  median metallicity-stellar 
mass relationship (blue dashed lines) of galaxies in the Illustris simulations.}
\label{fig3}
\end{figure}

We now explore the  accreted metallicity-stellar mass relationship for the 
central  galaxies in the Illustris simulations. In Figure \ref{fig3}, we  
plot the median metallicity of the total accreted stellar component of the 
Illustris galaxies as a  function of the accreted stellar mass colour-coded  
by the DM halo  mass of the galaxy ($11.0\,<\,\log_{10}\,\mathrm{M_{DM}}\,<\,14.0$). 
Illustris shows a  tight relationship between the metallicity and the stellar mass 
of the accreted component extending over three orders of magnitude 
in accreted stellar mass. 

The accreted metallicity-stellar mass relationship follows closely the Illustris
median metallicity-stellar mass relationship for galaxies, with the accreted 
stellar metallicity being on an average 0.3 dex lower than the total median 
metallicity (we discuss the colour-coding by dark matter halo mass later in 
\S \ref{sec:MW}). At higher accreted stellar mass ($\log\,\mathrm{M_{acc}} \ge 11.0$), 
the relationship between the metallicity and the accreted stellar mass 
tapers off.  Since the shape of the accreted metallicity-stellar mass relationship 
follows closely the shape of the metallicity-stellar mass relationship, it suggests
that the former is being driven by the accreted satellites following the galaxy
metallicity-stellar mass relationship.

\begin{figure}
\centering
\includegraphics[width=0.5 \textwidth,trim={0 1.5cm 0 0},clip]{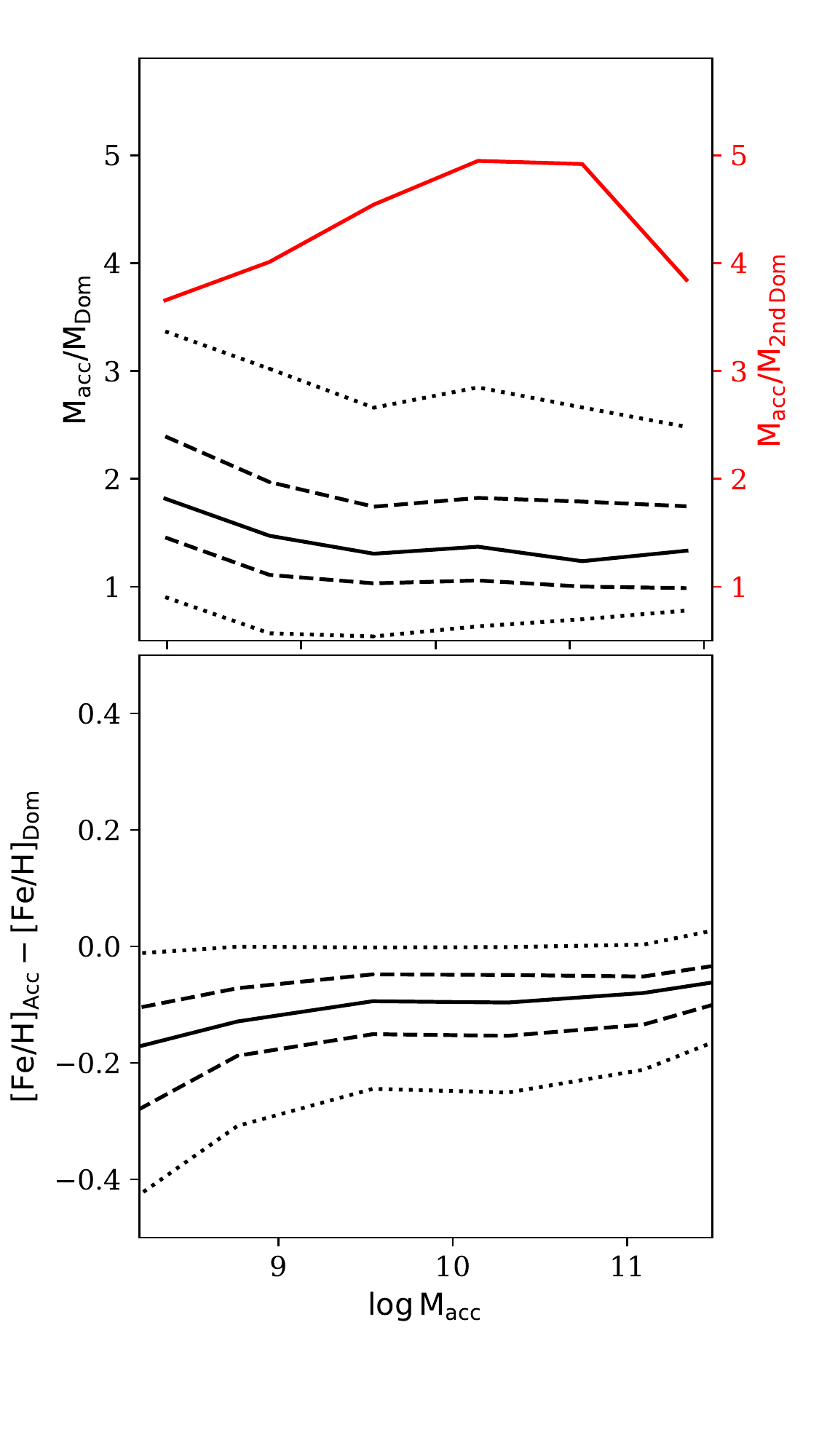}
\caption{{\bf Top panel}: The ratio of the accreted stellar mass to the stellar mass of the 
most massive accreted (``dominant'') progenitor as a function of accreted stellar mass. The 
black solid line indicates the median of the distribution. The dashed lines indicate the 25th 
and 75th percentiles. The dotted lines indicate the 5th and 95th percentiles. The red solid line 
indicates the median of the ratio of the accreted stellar mass to the stellar mass of the 
second most massive progenitor. The bulk of the accreted stellar component is contributed by
the dominant progenitor. {\bf Bottom panel}: The difference between the median metallicity  
of the accreted stellar component and the dominant progenitor as a function of accreted 
stellar mass and halo mass. } 
\label{fig7}
\end{figure}

In fact, the overall accreted metallicity-stellar mass relationship is being 
driven by the most massive accreted  (``dominant'') progenitor of the galaxy, 
which contributed the bulk of the accreted stellar material of the galaxy. We
demonstrate this in Figure \ref{fig7} by plotting the ratio of the  accreted 
stellar mass to the stellar mass of the dominant progenitor as a function of the  
accreted stellar mass. The median of this ratio is between 1 and 2.  In contrast, 
the ratio of the accreted stellar mass to the stellar mass of the second most 
massive progenitor is greater than 4. This suggests that the dominant accreted 
progenitor is on average twice as massive as the second most massive accreted 
progenitor, and  that the overall accreted stellar mass is dominated by the most 
massive progenitor.

The scatter in the ratio of the the  accreted stellar mass to the stellar mass of 
the dominant progenitor (noted by the 5th and 95th percentile of the distribution)
as a function of the accreted stellar mass indicates that a small fraction of galaxies 
are built up through small yet significant mergers. 

We  also plot the difference in metallicity between the accreted stellar component and 
the  dominant progenitor as a function of accreted stellar mass. This difference is 
nearly constant with accreted stellar mass and is around 0.1 dex in the expected sense 
that the dominant satellite is typically slightly more metal rich than the overall 
accreted halo. The maximum scatter in the  relationship is around 0.2\,dex. The low
amount of scatter indicates that the metallicity of the  dominant progenitor sets 
the accreted stellar metallicity of the galaxy and drives the general accreted 
metallicity-stellar mass relationship.

\section{Scatter in the Accreted Metallicity-Stellar Mass Relationship at Fixed 
Accreted Stellar Mass}
\label{sec:scatter}
We now turn our attention to understand what drives the scatter in the Illustris 
accreted-stellar mass relationship and how informative it is about the accretion history 
of a galaxy. As seen in Figure \ref{fig3}, there is 0.2 dex scatter at a given accreted 
stellar mass in the Illustris accreted metallicity-stellar mass relationship. The 
scatter at a fixed accreted stellar mass reflects the different possible accretion 
histories which can build up an accreted stellar component of that mass.

\begin{figure}
\centering
\includegraphics[width=0.5\textwidth]{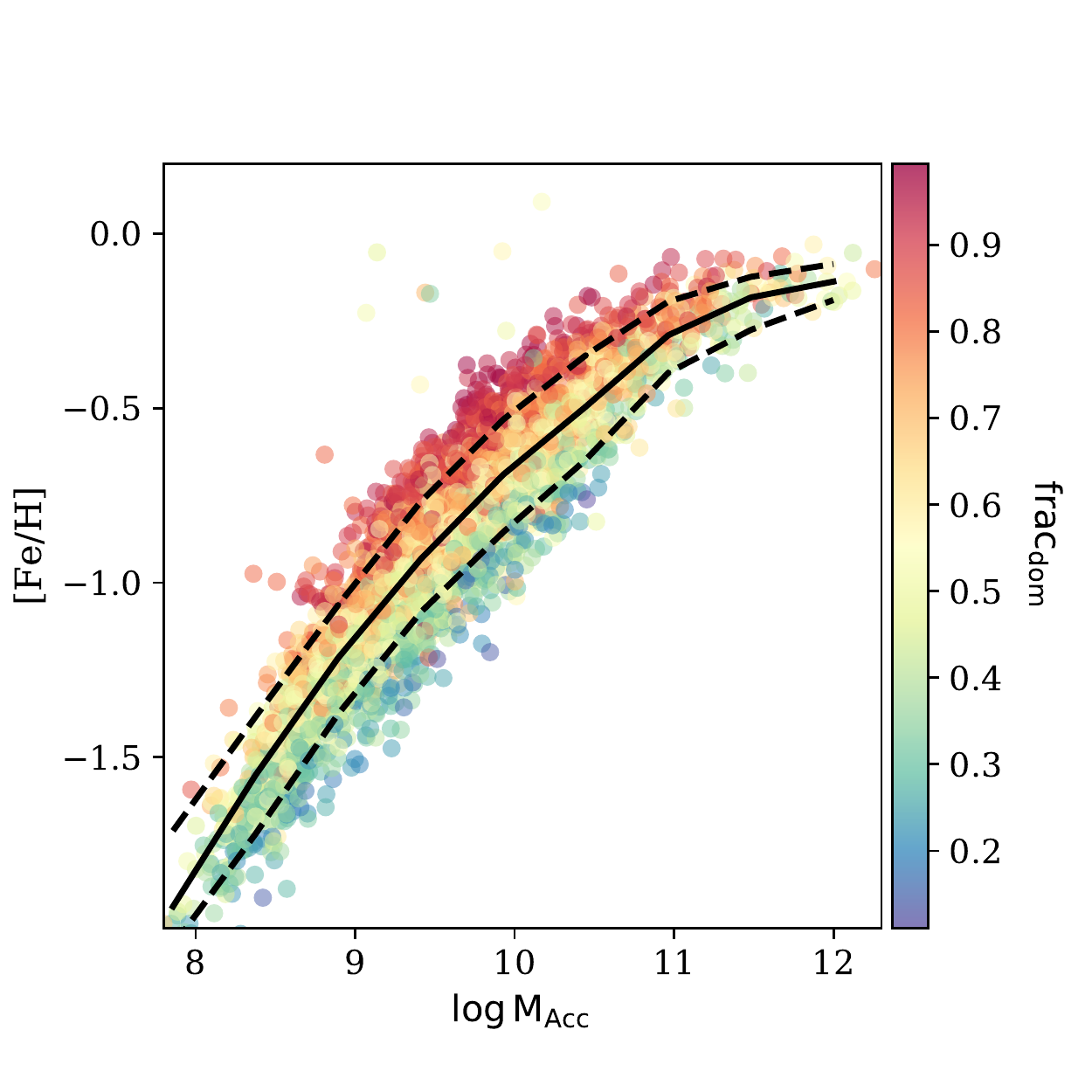}
\caption{Accreted metallicity-stellar mass relationship colour coded by the 
fraction of accreted stellar mass ($\mathrm{frac_{Dom}}$) contributed 
by the dominant progenitor. The scatter in the accreted metallicity-stellar mass
relationship encodes information about $\mathrm{frac_{Dom}}$.} 
\label{fig4}
\end{figure}

We parametrize the accretion history of a galaxy in terms of the fraction of accreted 
stellar material contributed by the dominant progenitor $\mathrm{frac_{Dom}}$. 
This allows us to study how the accreted stellar metallicity varies as a function of the 
most massive accretion event of the galaxy. In  Figure \ref{fig4}, we consider 
the accreted metallicity-stellar mass relationship colour coded by $\mathrm{frac_{Dom}}$. 
At a  given accreted stellar mass, we find that galaxies with a higher accreted 
stellar metallicity have their accreted haloes built up predominately by a 
single merger event, while galaxies with a lower accreted stellar  metallicity have 
their haloes built up through a number of accreted  satellites. 

\begin{figure}
\centering
\includegraphics[width=0.5\textwidth]{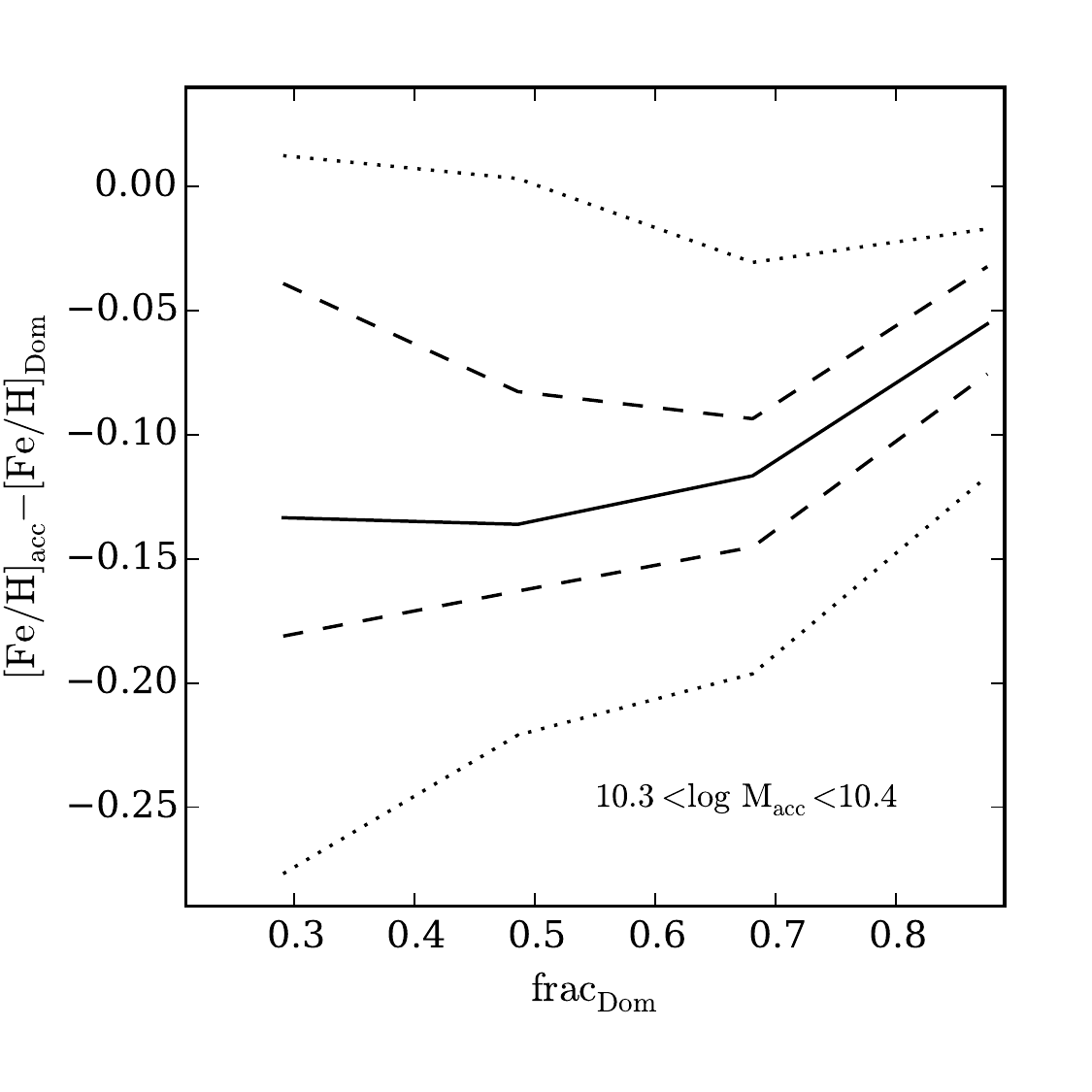}
\caption{The difference in accreted stellar metallicity and the metallicity of the 
dominant progenitor as a function of $\mathrm{frac_{Dom}}$ for galaxies at a fixed 
accreted stellar mass ($10.3 < \log\, \mathrm{M_{acc}}  < 10.4$). The dashed lines 
indicate the 25th and 75th percentiles. The dotted lines indicate the 5th and 95th 
percentiles. At a fixed accreted stellar mass, the spread in accreted metallicity 
is driven by the most dominant progenitor.} 
\label{fig4b}
\end{figure}

While Fig.\ \ref{fig4} shows that high metallicity halos for a given accreted mass 
tend to have a higher $\mathrm{frac_{Dom}}$, it is interesting to consider more 
closely the contribution of the dominant progenitor to the metallicity 
at a fixed accreted stellar mass ($10.3 < \log\, \mathrm{M_{acc}}  < 10.4$) 
in  Figure \ref{fig4b}. At higher $\mathrm{frac_{Dom}}$, the metallicity 
of the accreted stellar component approaches the metallicity of the dominant 
progenitor, with a scatter considerably smaller than 0.1\,dex.  At lower 
$\mathrm{frac_{Dom}}$, the median metallicity of the 
accreted stellar component is about 0.15 dex lower than the metallicity of 
the dominant progenitor.  Although there is an increase in scatter in the 
contribution of the dominant  progenitor to the metallicity of the accreted 
stellar halo at lower  $\mathrm{frac_{Dom}}$, the scatter is $<$0.1\,dex, 
considerably smaller than the $\sim 0.2$\,dex scatter of the accreted 
metallicity--mass relation. This is consistent with the fact that at 
lower $\mathrm{frac_{Dom}}$, a number of  progenitors contribute to the 
build up the accreted stellar component, leading to an increase in scatter 
in the accreted metallicity-stellar mass relationship. From Figures \ref{fig4} 
and \ref{fig4b}, we conclude that while there are a range of possible accretion 
paths to reach a given accreted stellar mass, the metallicity of the
accreted stellar component is largely set by the dominant progenitor.

\begin{figure}
\centering
\includegraphics[width=0.5\textwidth]{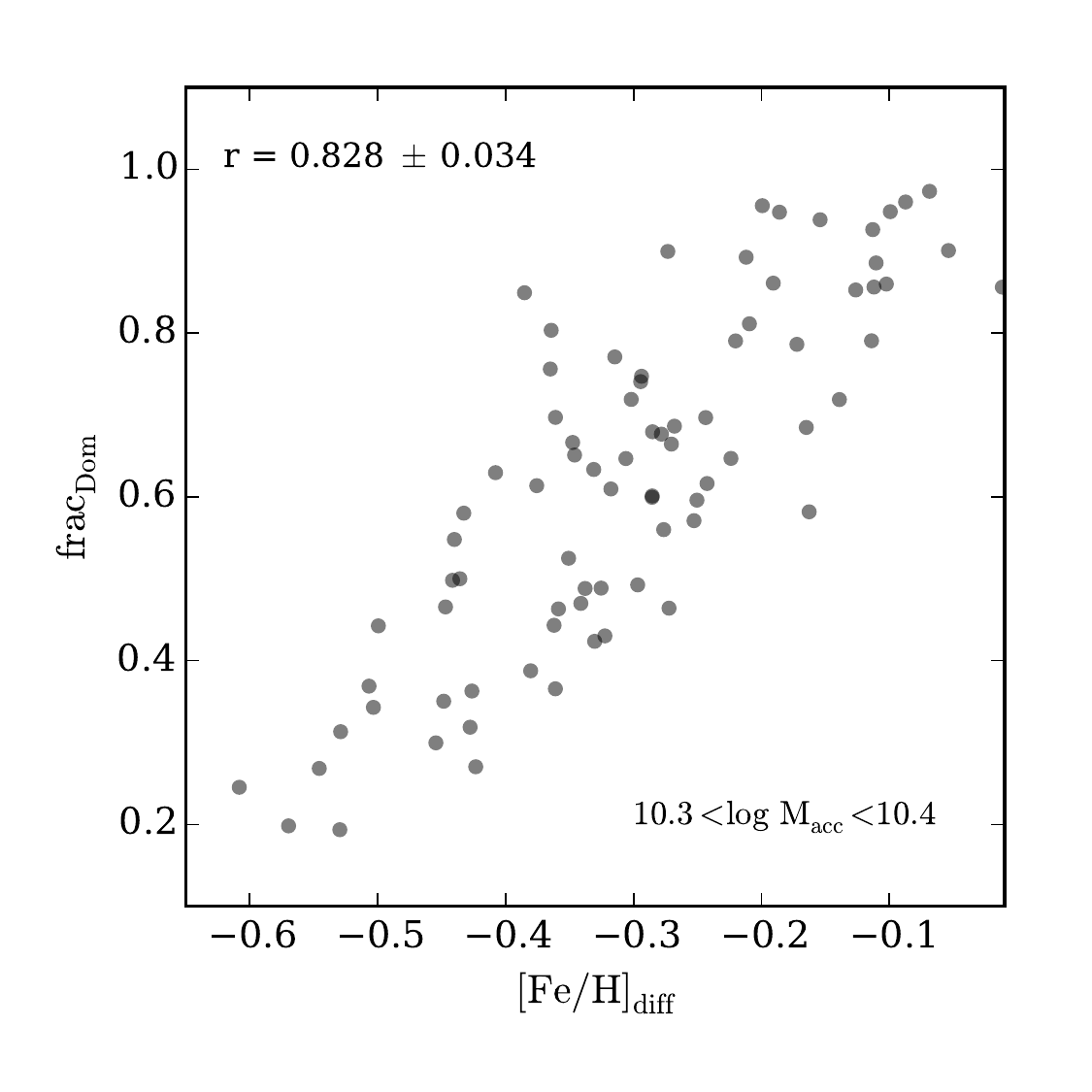}
\caption{The fraction of accreted stellar material contributed by the dominant 
progenitor $\mathrm{frac_{Dom}}$ as a function of 
$\mathrm{[Fe/H]_{diff}} = \mathrm{[Fe/H]_{acc} - [Fe/H]_{acc\,predict}}$ at a fixed
accreted stellar mass ($10.3 < \log\, \mathrm{M_{acc}}  < 10.4$). The difference in accreted stellar metallicity ($\mathrm{[Fe/H]_{diff}}$) encodes information about $\mathrm{frac_{Dom}}$. } 
\label{fig4c}
\end{figure}

The scatter in  accreted metallicity  can inform us about the mass of the 
dominant progenitor. If the accreted stellar component was built from a single
massive progenitor, the  metallicity of the accreted stellar component would
be given by the Illustris galaxy metallicity--stellar mass relationship.  
This motivates us to define the  quantity $\mathrm{[Fe/H]_{diff}}$, i.e., 
the difference between the median accreted metallicity and the metallicity 
of the accreted stellar component if it is built up from a single massive progenitor 
($\mathrm{[Fe/H]_{diff}} = \mathrm{[Fe/H]_{acc} - [Fe/H]_{acc\,predict}}$),
where  $\mathrm{ [Fe/H]_{acc\,predict}}$ is the stellar metallicity assuming
the accreted stellar component was built from a single massive progenitor.
We estimate  $\mathrm{ [Fe/H]_{acc\,predict}}$ using the Illustris galaxy 
metallicity-stellar mass relationship, assuming that  $\mathrm{M_{*}} \sim 
\mathrm{M_{acc}}$. Given the {\it ansatz} that the metallicity of an 
accreted galaxy only depends upon its stellar mass, this quantity 
$\mathrm{[Fe/H]_{diff}}$ allows  us to quantify $\mathrm{frac_{Dom}}$ 
in terms of metallicity. In Figure \ref{fig4c}, we plot $\mathrm{frac_{Dom}}$
as a function of  $\mathrm{[Fe/H]_{diff}}$ at a fixed accreted stellar 
mass ($10.3 < \log\, \mathrm{M_{acc}}  < 10.4$). There is a strong 
correlation between  $\mathrm{frac_{Dom}}$ and $\mathrm{[Fe/H]_{diff}}$, 
allowing us to predict the  mass of the dominant progenitor at a given
accreted stellar mass.

\section{Scatter in the Accreted Metallicity-Stellar Mass Relationship at Fixed 
Virial Size}
\label{sec:MW}
At a fixed accreted stellar mass, galaxies span a range in virial masses and 
hence radii (see Figure \ref{fig3}). In order to explore the scatter in the  
accreted-metallicity stellar mass relationship, it is imperative to control 
for galaxies of similar sizes. We do this in this section by comparing 
galaxies in a narrow DM halo mass range. This allows connections to be made 
between our work and that of e.g., \citet{Bullock2005}, \citet{Deason2016}, 
\citet{Amorisco2017} or Monachesi et al.\ (in preparation).  As seen in 
Figure \ref{fig3}, galaxies belonging to a narrow DM halo mass range --- such 
as MW-like mass galaxies --- occupy a narrower and steeper locus on the 
accreted metallicity--stellar mass plane (illustrated by the fits to narrow 
DM halo mass ranges, shown as thin solid lines, in Fig.\ \ref{fig3}). We note 
that similar steeper relationships between accreted metallicity  and stellar  
mass of the halo are found if one considers narrow ranges in stellar mass  or velocity dispersion. 

\begin{figure}
\centering
\includegraphics[width=0.5\textwidth]{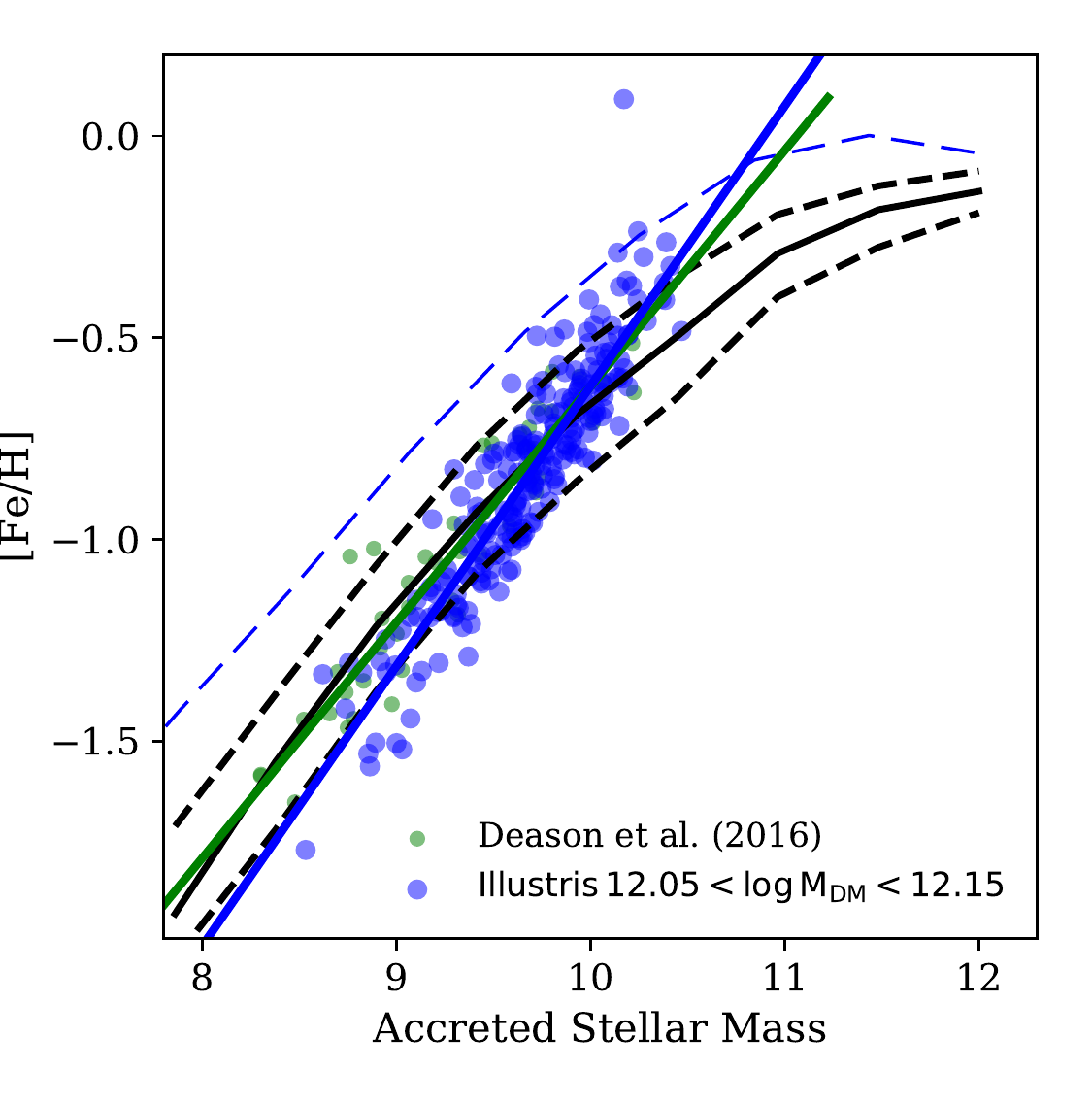}
\caption{Comparison of the 45 haloes from the \protect\cite{Deason2016} models and 
Illustris galaxies in the narrow DM halo mass range ($12.05 \le 
\mathrm{\log\,M_{DM}} \le 12.15$).  The solid blue and green lines indicate linear
fits to the Illustris and \protect\citeauthor{Deason2016} models respectively.
The red solid and dashed lines indicate the median and the  16/84th percentile of 
the parent distribution of  central Illustris galaxies as shown in Figure \ref{fig3}. 
The blue dashed line indicates the median metallicity-stellar mass relationship of 
galaxies in the Illustris simulations.} 
\label{fig4d}
\end{figure}

For the purposes of this study, we examine MW-like mass galaxies in DM halo mass 
range ($12.05 \le \mathrm{\log\,M_{DM}} \le 12.15$). Before we begin comparing 
the physical properties of MW-like mass galaxies, we examine the  accreted 
metallicity-stellar mass relationship  for these galaxies. In Figure \ref{fig4d}, 
we plot the accreted metallicity-stellar mass relationship of the Illustris galaxies  
for MW-like mass galaxies in blue.  We note that the slope of the accreted 
metallicity-stellar mass relationship for MW-mass like galaxies is much steeper 
than the Illustris metallicity-stellar mass relationship for all galaxies (blue dashed 
line).

We also compare these galaxies with the metallicity of total accreted stellar component of the 
45 model stellar haloes from \cite{Deason2016} in green. We find that while the results agree 
in overall normalisation, they have a small difference in slopes with the Illustris 
galaxies producing a steeper accreted metallicity-stellar mass relationship than 
\citeauthor{Deason2016}. This is a direct consequence of the difference in the input stellar mass--halo mass and 
metallicity--stellar mass relationships between the Illustris simulations and \cite{Deason2016}.  
In the latter, the stellar mass is assigned to DM haloes with the relationship published in 
\cite{Garrison-Kimmel2014}, while  the metallicity assigned to the stellar particles follows \cite{Kirby2013} 
with explicit time-evolution derived from the simulations of \cite{Ma2016}. On 
the other hand, the Illustris simulations has a much steeper galaxy 
metallicity--stellar mass relationship with hardly any time evolution. 

\begin{figure}
\centering
\includegraphics[width=0.5\textwidth]{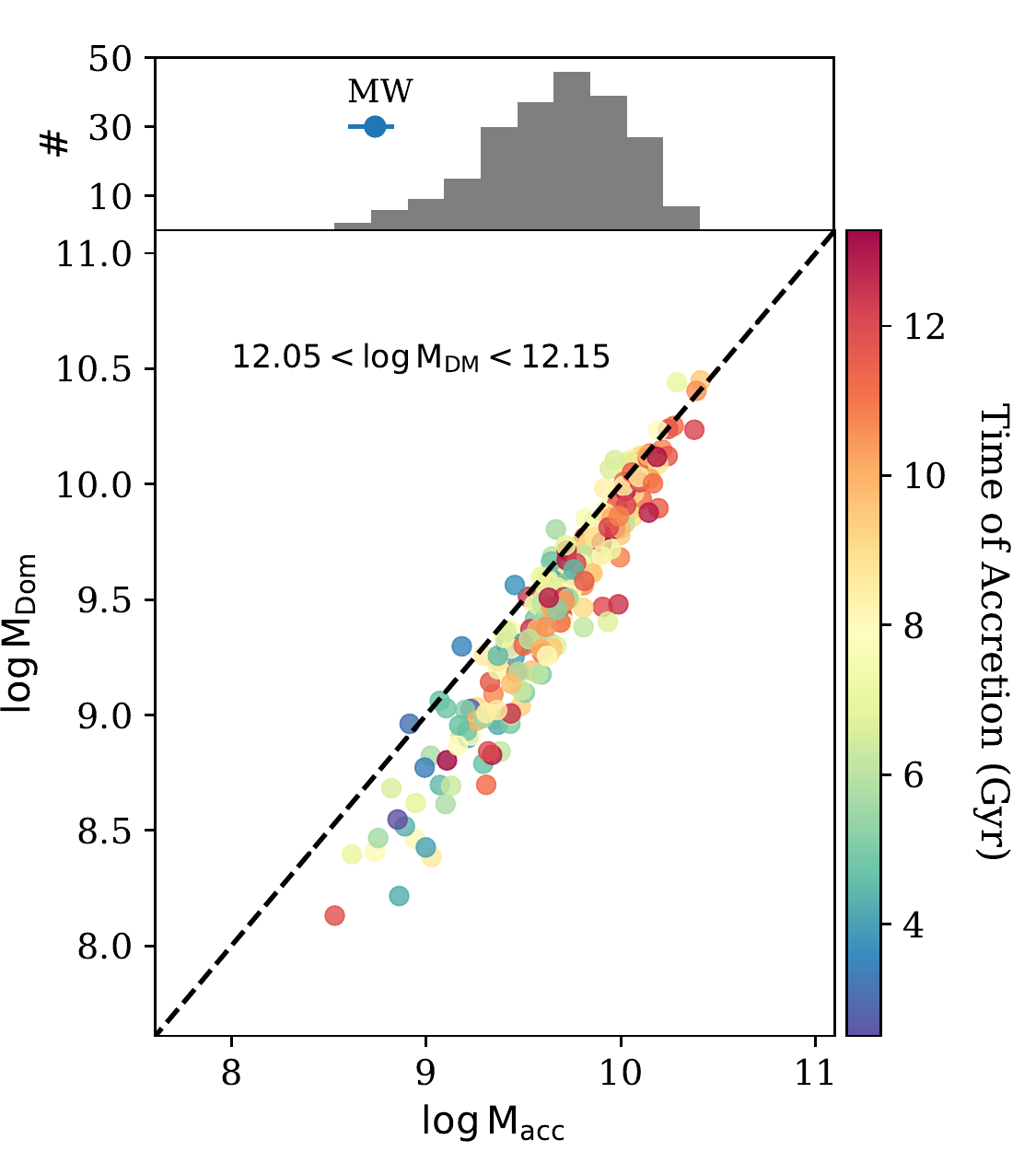}
\caption{{\bf Top panel}: The distribution of accreted stellar masses 
for galaxies with DM halo masses, $12.05 \le \mathrm{\log\,M_{DM}} 
\le 12.15$.  The accreted stellar mass of the MW is indicated (Bland-Hawthorn \& Gerhard 2016, following Bell et al. 2008) and is at the lower end of the distribution of MW-mass galaxies, hinting to a quiet accretion history. {\bf Bottom panel}: The stellar mass of the dominant progenitor as a function of the accreted stellar mass, 
colour coded by the time of accretion of the progenitor, where red signifies 
early accretions at high redshift and blue signifies a dominant accretion close 
to the present day. The dashed line shows $\mathrm{frac_{Dom}} = 1$; galaxies 
below this line have $\mathrm{frac_{Dom}}<1$. In general, larger progenitors are accreted much later in time.} 
\label{fig8}
\end{figure}

In a narrow DM halo mass range, the  accreted stellar components of these galaxies
show a rich diversity in accreted stellar mass.  The top panel of Figure 
\ref{fig8} shows the  distribution of accreted stellar masses for  MW-like 
mass galaxies. We note that the measured accreted stellar mass of the MW puts 
it at the tail end of the distribution of accreted stellar mass, highlighting its
unusual accretion history.

The bottom panel of Figure \ref{fig8} quantifies the relationship between the 
stellar mass of the  dominant progenitor to the accreted stellar mass for 
MW-like mass galaxies, colour-coded with the time  of accretion of the dominant 
progenitor. The dashed line shows $\mathrm{frac_{Dom}} = 1$; galaxies below this 
line have $\mathrm{frac_{Dom}}<1$. Galaxies with smaller accreted stellar mass
have significantly smaller dominant  progenitors (smaller $\mathrm{frac_{Dom}}$), 
consistent with the idea that their  accreted stellar component was built up 
through a number of accretion events. Given galaxies of a similar DM halo mass, 
there exists a scattered correlation between $\mathrm{frac_{Dom}}$ and the 
accreted  stellar mass with a significant scatter, consistent with the results 
of \cite{Deason2016} and \cite{Amorisco2017b}.

For galaxies of similar DM halo mass, the dominant progenitors of galaxies
with lower $\mathrm{frac_{Dom}}$ are accreted earlier in time, while larger 
dominant progenitors are accreted later in  time (as seen in Figure \ref{fig8}). 
While there exists a correlation between the mass of the dominant progenitor
and the time of its accretion, there is a considerable scatter in the relationship.

Additional information is needed to break the degeneracy between $\mathrm{frac_{Dom}}$  
and the  time of accretion in order to infer the stellar mass and the time of accretion 
of the dominant progenitor. This information may be contained in the other physical 
properties of the accreted stellar component (including its morphology), to which we turn 
to in Section \ref{sec:diversity}.

\begin{figure}
\centering
\includegraphics[width= 0.45 \textwidth]{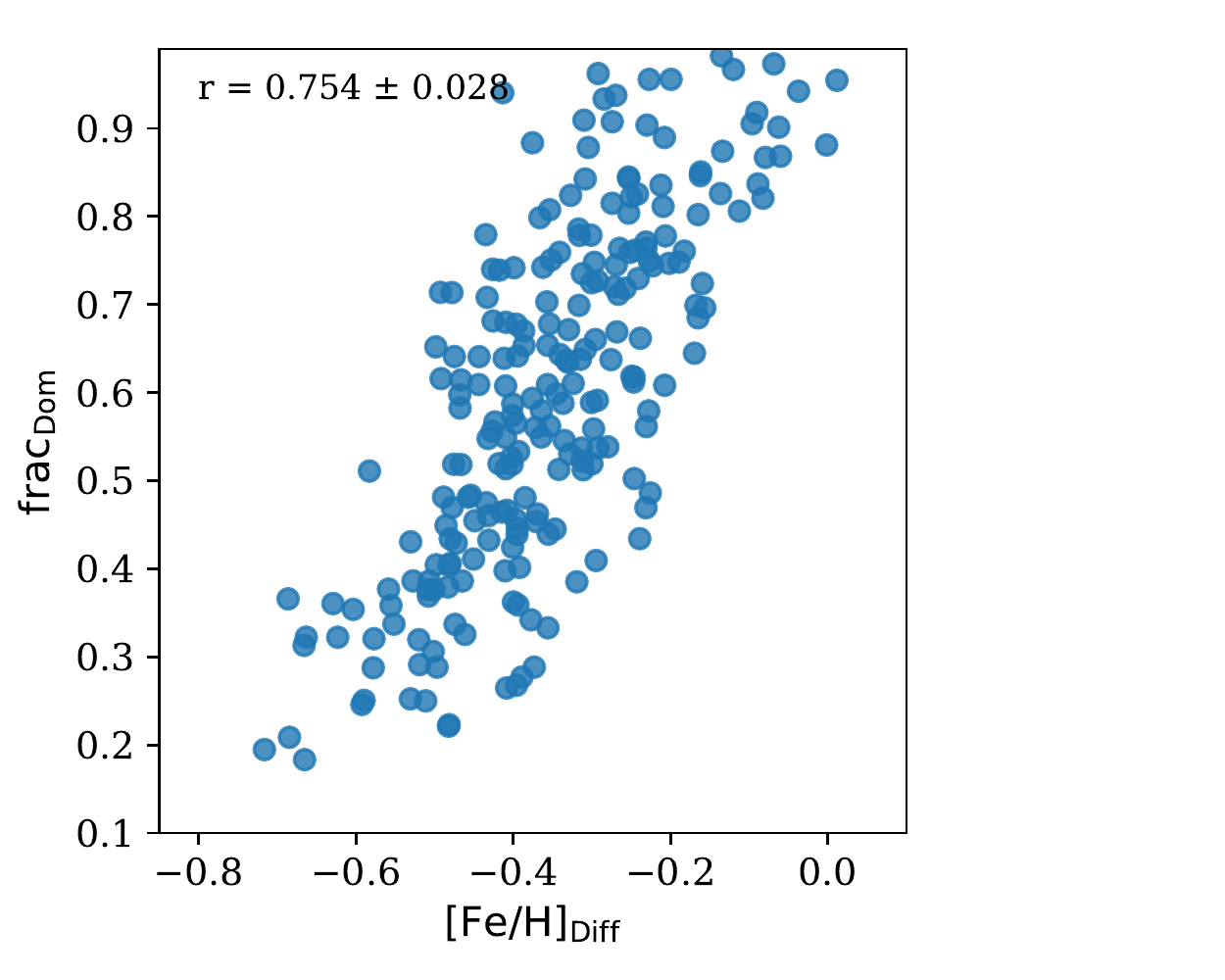}
\includegraphics[width= 0.45 \textwidth]{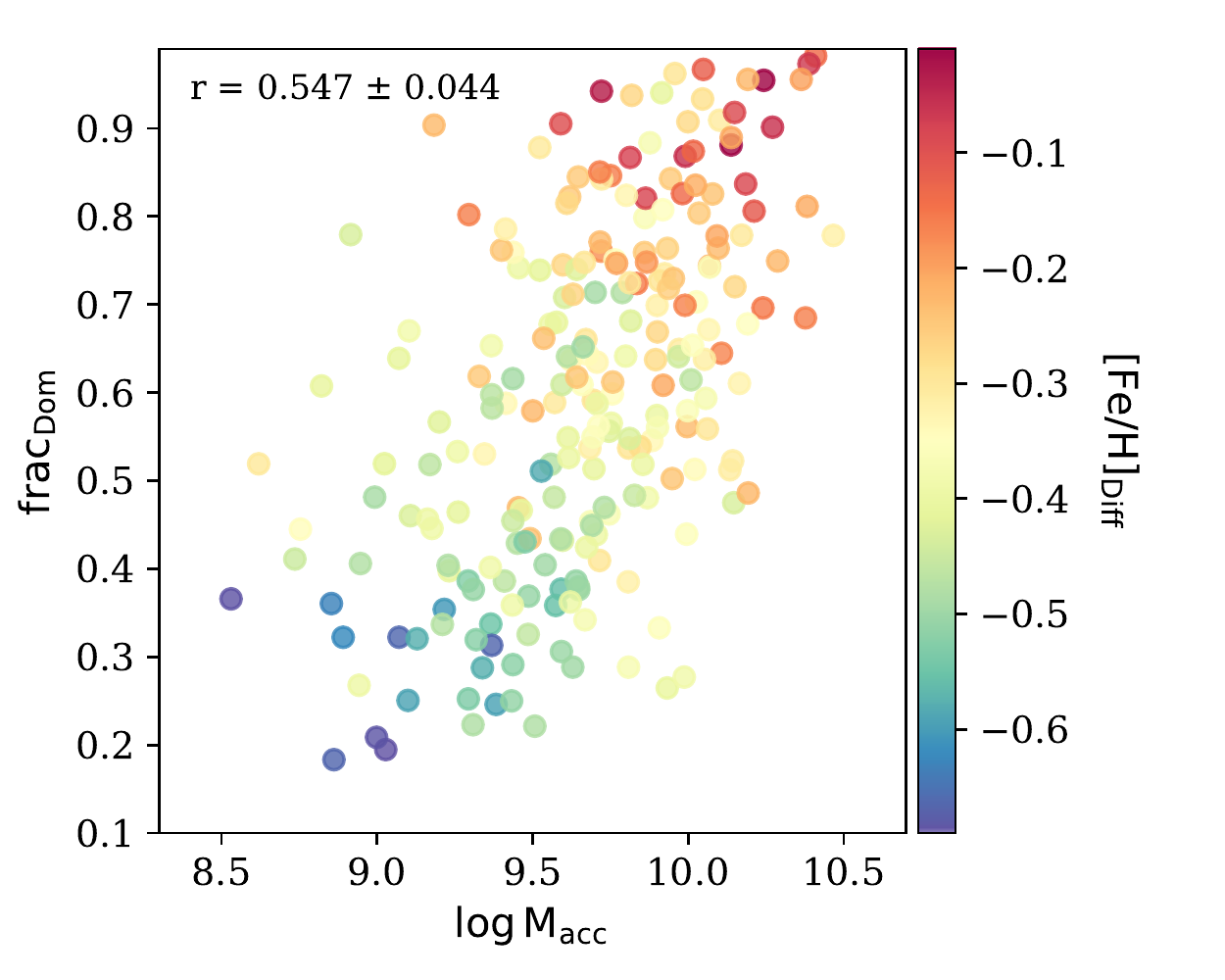}
\caption{{\bf Top Panel}: $\mathrm{frac_{Dom}}$ as a function of $\mathrm{[Fe/H]_{diff}}$ for MW-like mass galaxies. {\bf Bottom Panel}: $\mathrm{frac_{Dom}}$ as a function of the accreted stellar mass colour-coded 
by $\mathrm{[Fe/H]_{diff}}$ for MW-like mass galaxies.
We also calculate the Pearson Correlation coefficient (r) for the two 
relationships. The best predictor of $\mathrm{frac_{Dom}}$ of 
MW-like mass galaxies is the accreted stellar metallicity 
at a fixed accreted stellar mass.} 
\label{fig_metals_diff}
\end{figure}

Even with the additional constraint of the DM halo mass of the galaxy, it is 
instructive to ask which quantity, the accreted stellar mass or the accreted 
stellar metallicity, is most informative about $\mathrm{frac_{Dom}}$. In Figure 
\ref{fig_metals_diff}, we plot $\mathrm{frac_{Dom}}$ as a function of the  accreted 
stellar mass and $\mathrm{[Fe/H]_{diff}}$. From the Pearson correlation coefficients, 
we confirm that the accreted stellar metallicity (or $\mathrm{[Fe/H]_{diff}}$) is more 
informative about  $\mathrm{frac_{Dom}}$ than the accreted stellar mass 
for MW-like mass galaxies. This is consistent with the results found in 
Section \ref{sec:scatter} at a fixed accreted stellar mass.

To set further constraints on $\mathrm{frac_{Dom}}$, it is best to compare 
MW-like mass galaxies at a fixed accreted stellar mass. We do this in the 
right panel of  Figure \ref{fig_metals_diff}, where we plot $\mathrm{frac_{Dom}}$ 
as a function  of the  accreted stellar mass colour-coded by $\mathrm{[Fe/H]_{diff}}$. We see that at a given accreted stellar mass,  $\mathrm{frac_{Dom}}$ correlates with $\mathrm{[Fe/H]_{diff}}$ allowing us to distinguish galaxies according to the mass of their dominant progenitor. The best predictor of $\mathrm{frac_{Dom}}$ is the accreted stellar metallicity at a fixed accreted stellar mass.

\begin{figure}
\centering
\includegraphics[width=0.5 \textwidth]{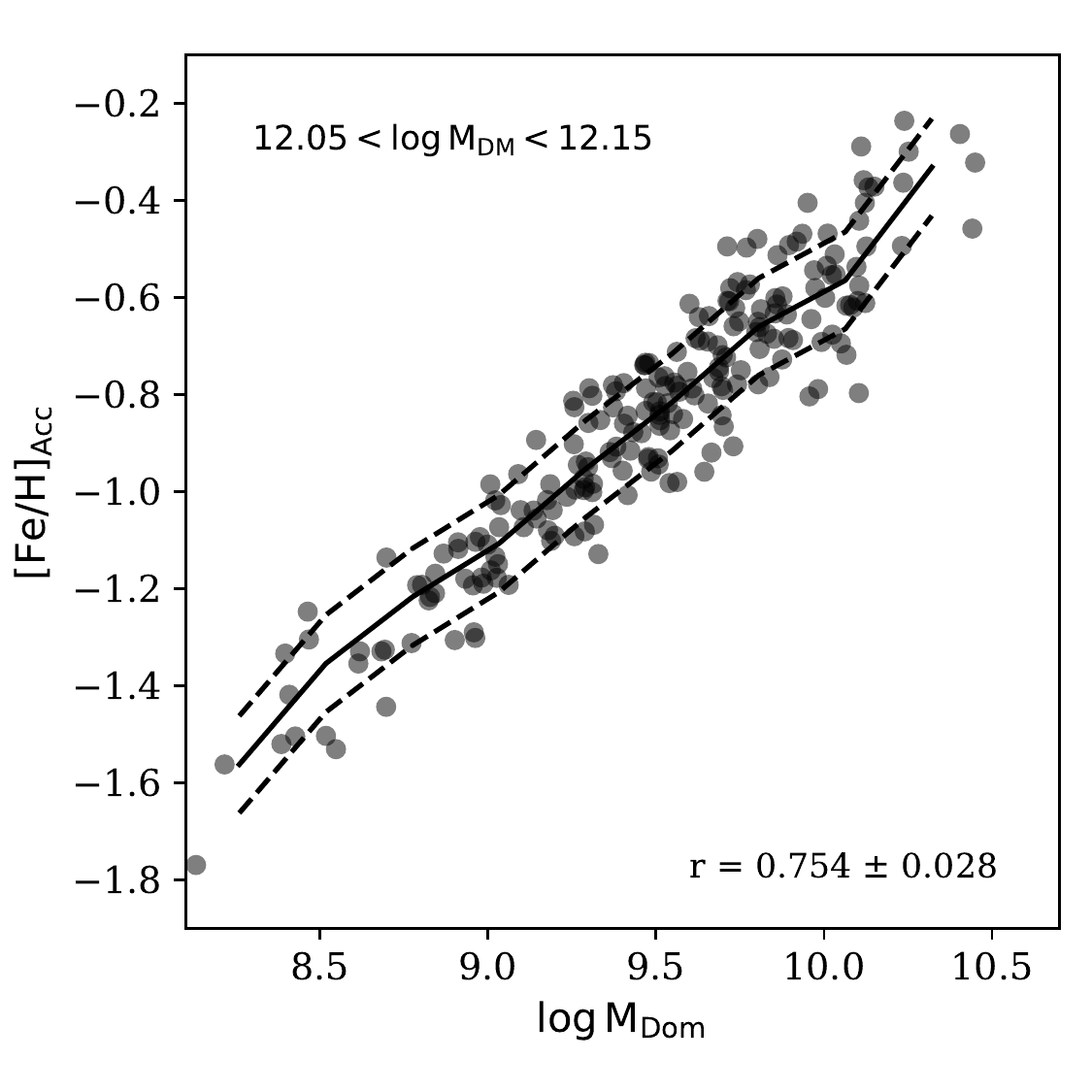}
\caption{The relationship between the accreted stellar metallicity and the mass of the dominant progenitor for MW-like mass galaxies. The relationship follows the Illustris galaxy metallicity-stellar mass relationship seen in Figure \ref{fig1}. The dashed lines indicate the 1-$\sigma$ scatter in the relationship. We calculate the Pearson correlation coefficient (r) of the relationship. The accreted stellar metallicity is a strong predictor of the stellar mass of the dominant progenitor.}
\label{fig21}
\end{figure}

It is important to note that while $\mathrm{[Fe/H]_{diff}}$ correlates best with  
$\mathrm{frac_{Dom}}$, the accreted stellar metallicity  ($\mathrm{[Fe/H]_{Acc}}$) 
is the best predictor of $\mathrm{M_{Dom}}$, through the galaxy metallicity-stellar 
mass relationship (See Figure \ref{fig21}). The scatter in the relationship between
$\mathrm{[Fe/H]_{Acc}}$ and $\mathrm{M_{Dom}}$ is the same as the scatter in the 
Illustris galaxy metallicity-stellar mass relationship.

\subsection{Diversity of physical properties of the accreted stellar component of MW-like mass galaxies}
\label{sec:diversity}
Due to dynamical friction and tidal stripping, the accreted stellar component of 
MW-like mass galaxies display a large diversity in their physical morphology.  
The influence of dynamical friction on the deposition of tidally shredded material
from incoming satellites in the accreted stellar component of a galaxy have been 
explored in depth by \cite{Amorisco2017} using idealised N-Body simulations and 
\cite{Rodriguez-Gomez2016} for the Illustris galaxies. In general, the deposition 
of accreted stellar material within the hosts depends upon the mass of the 
progenitor, its compactness and the time of its accretion. More massive 
progenitors deposit their stellar material at smaller  radii deep within the host, 
while less massive progenitors deposit their stellar material at  the outskirts of 
the galaxies. Satellites accreted at a higher redshift deposit their  stellar  
material at smaller radii within the host, since the hosts were physically 
smaller at that time. More concentrated progenitors deposit their material 
at smaller radii. In particular, \cite{Amorisco2017} showed that due to the 
mass-concentration scatter in DM haloes, satellites that are  1$\sigma$ more 
concentrated than average can deposit their stars at radii that are closer 
in by a factor of $\sim$ 2.5 in mass  (and average concentration).  It would 
be difficult to disentangle the mass of the  progenitor or the time of its 
accretion purely from spatial information  of the accreted stellar component.

\begin{figure*}
\centering
\includegraphics[width=\textwidth]{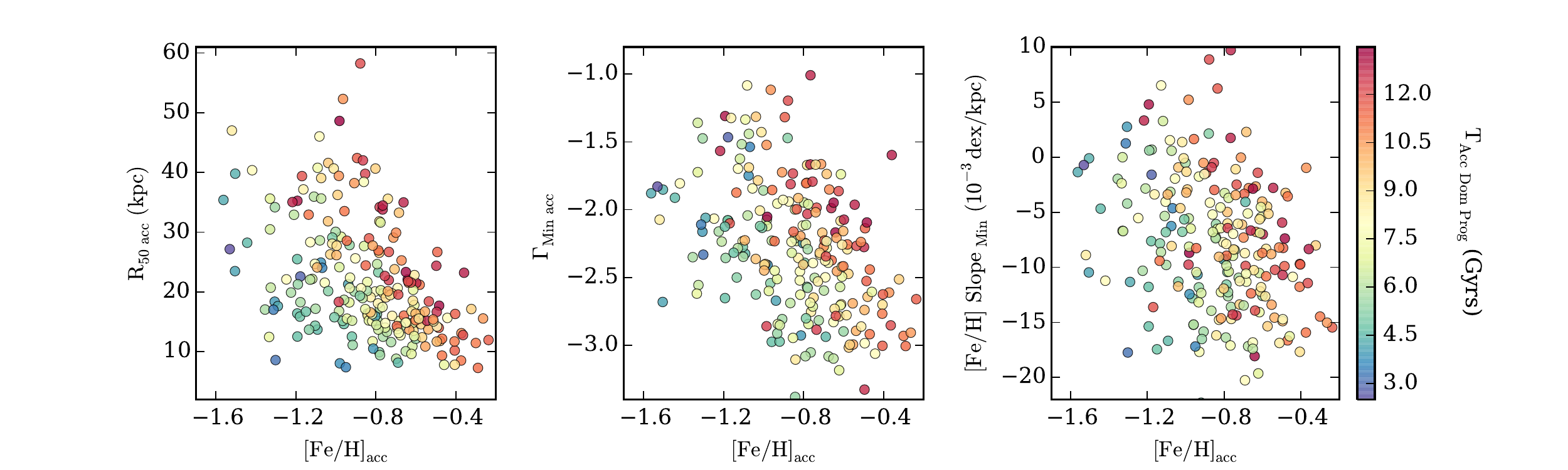}
\caption{The diversity of the accreted stellar component of MW-like mass galaxies: The 3D 
half-mass radius, the  power-law density slope and the metallicity gradient along the minor 
axis of the accreted stellar component as a function of the accreted stellar metallicity 
for MW-like  mass galaxies. The galaxies are colour-coded by the time of accretion of the
dominant progenitor, where red signifies early accretions at high redshift and 
blue signifies a dominant accretion close to the present day. The physical size, the surface mass
density slope as well as the metallicity gradient encodes information about the time of accretion
of the dominant progenitor.} 
\label{fig_MW_diversity}
\end{figure*}

In  Figure \ref{fig_MW_diversity}, we demonstrate the diversity in the physical 
properties of the accreted stellar component for MW-like mass galaxies found in the 
Illustris simulations.  In particular, we  show the 3D half-mass radius 
($\mathrm{R_{50\,acc}}$), the power-law density slope ($\mathrm{\Gamma_{min\,acc}}$) 
and the metallicity gradient along  the minor axis of the accreted stellar component.
We plot these quantities as a function of the accreted stellar metallicity of the galaxy. 

First of all, we find a rich diversity in the physical morphology of the accreted
stellar component. Galaxies with a larger accreted stellar metallicity tend to have more 
compact accreted stellar component (smaller $\mathrm{R_{50\,acc}}$) than galaxies with  
a lower accreted stellar metallicity.  Similarly, both the power-law density slope 
and the slope of the accreted metallicity gradient becomes steeper with increasing 
accreted stellar metallicity. The decreasing trend in all these relationships with 
increasing accreted stellar metallicity reflects the correlation of increasing 
accreted stellar metallicity with the mass of the dominant progenitor ($\mathrm{M_{Dom}}$; 
see Figure \ref{fig21}).  Galaxies with larger dominant progenitors ($\mathrm{M_{Dom}}$) 
tend to have more compact accreted stellar components.

Secondly, there is also a significant scatter in these quantities at fixed
accreted stellar metallicity. At lower accreted stellar metallicity, there 
is a substantial  spread in the physical size of the accreted stellar component. 
The scatter in the relationships decreases with increasing accreted stellar 
metallicity. This is because the physical properties of the accreted stellar
component are being dominated by a single large progenitor, due to increasing 
$\mathrm{frac_{Dom}}$ with increasing accreted stellar metallicity.

Finally, the scatter in these relationships at a fixed accreted stellar 
metallicity correlates with the time of accretion of the dominant 
progenitor satellite (see colour-coding in Figure \ref{fig_MW_diversity}). 
In general, satellites accreted  earlier in time deposit their stellar  
material at smaller radii within the host, because their hosts are much 
smaller. Concentrated progenitors also deposit their disrupted material 
at smaller radii. Moreover, the concentration of DM haloes reflects the 
matter density of the Universe at the time of their  formation \citep{Navarro1997}. 
Accreted DM haloes formed earlier tend to be more 
concentrated  and sink more towards the centre of the galaxy due to 
dynamical friction, than accreted DM haloes formed later which tend to be 
less concentrated and deposit their disrupted accreted material at large 
galacto-centric distances. This induces a further correlation with time, 
wherein satellites which are formed earlier deposit their accreted stellar 
material closer to the centre of the galaxy. The scatter in the 
physical ``spatial'' properties of the accreted stellar components of galaxies is 
informative about the time of its build-up: the physical size, the surface mass 
density slope as well as the metallicity gradient of the accreted stellar component 
encodes information about the time of accretion of the most dominant progenitor.

\begin{figure*}
\centering
\includegraphics[width=0.8\textwidth]{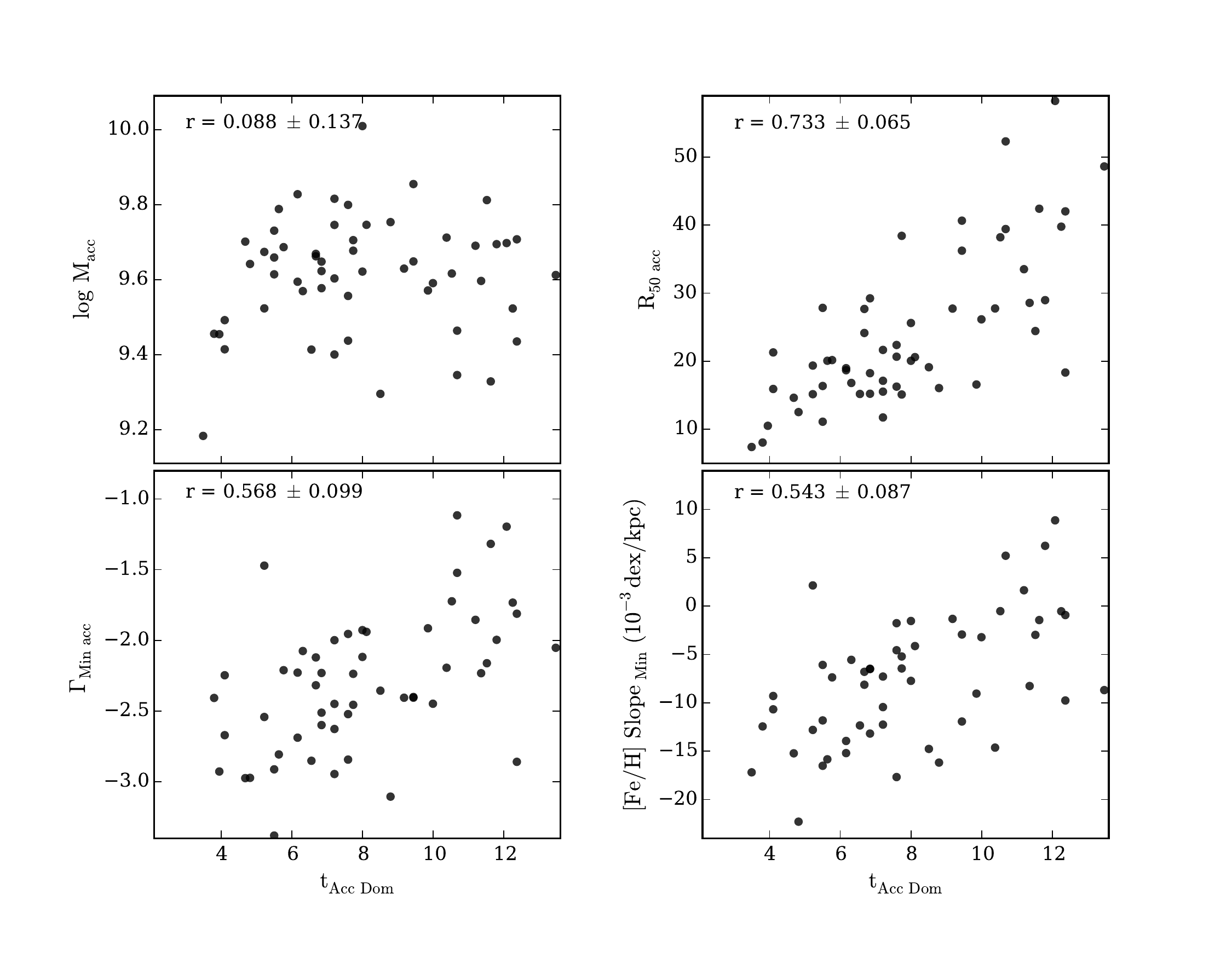}
\caption{For MW-like mass galaxies with accreted stellar metallicites 
($-1 < \mathrm{[Fe/H]_{acc}} < -0.8$), we plot the correlations 
between  the time of accretion of the dominant progenitor and 
$\mathrm{\log\,M_{acc}}$,  $\mathrm{R_{50\,acc}}$, $\mathrm{\Gamma_{acc\,min}}$ 
as well as the stellar metallicity gradient of the accreted component along 
the minor axis. We also calculate the Pearson Correlation coefficient (r) for the 
relationships. The physical size, the surface mass density slope as well as the metallicity
gradient of the accreted stellar component encodes information about the time of accretion 
of the most dominant progenitor, while the mass of the accreted stellar component contains 
no information about the time of accretion.} 
\label{fig_timefrac_predict}
\end{figure*}

We quantify this further in Figure \ref{fig_timefrac_predict} by choosing 
galaxies with the similar masses of the dominant progenitor. We do so by 
choosing galaxies using the more observationally accessible quantity 
$\mathrm{[Fe/H]_{acc}}$ (see Section \ref{sec:minor}) employing the  
correlation between $\mathrm{[Fe/H]_{acc}}$ and $\mathrm{M_{Dom}}$
(see Figure \ref{fig21}). In Figure \ref{fig_timefrac_predict}, we choose 
galaxies such that   $-1 < \mathrm{[Fe/H]_{acc}} < -0.8$. We find that  the 
3D half-mass radius correlates most with the time of accretion of the 
dominant progenitor. We also notice that the other indicators indicating 
spatial information, that is, the 2d power-law density slope and the 
metallicity gradients of  the  accreted  component along the minor axis 
also  encodes some information  about the  time of accretion. On the other 
hand, we find that accreted stellar mass is a poor predictor of the time 
of accretion consistent with Figure \ref{fig8}. We also note that the  
presence of time evolution in the galaxy metallicity-stellar mass relationship  
will further affect the metallicity gradients. This effect cannot be studied 
in the Illustris simulation.

\begin{figure*}
\centering
\includegraphics[width=0.8\textwidth]{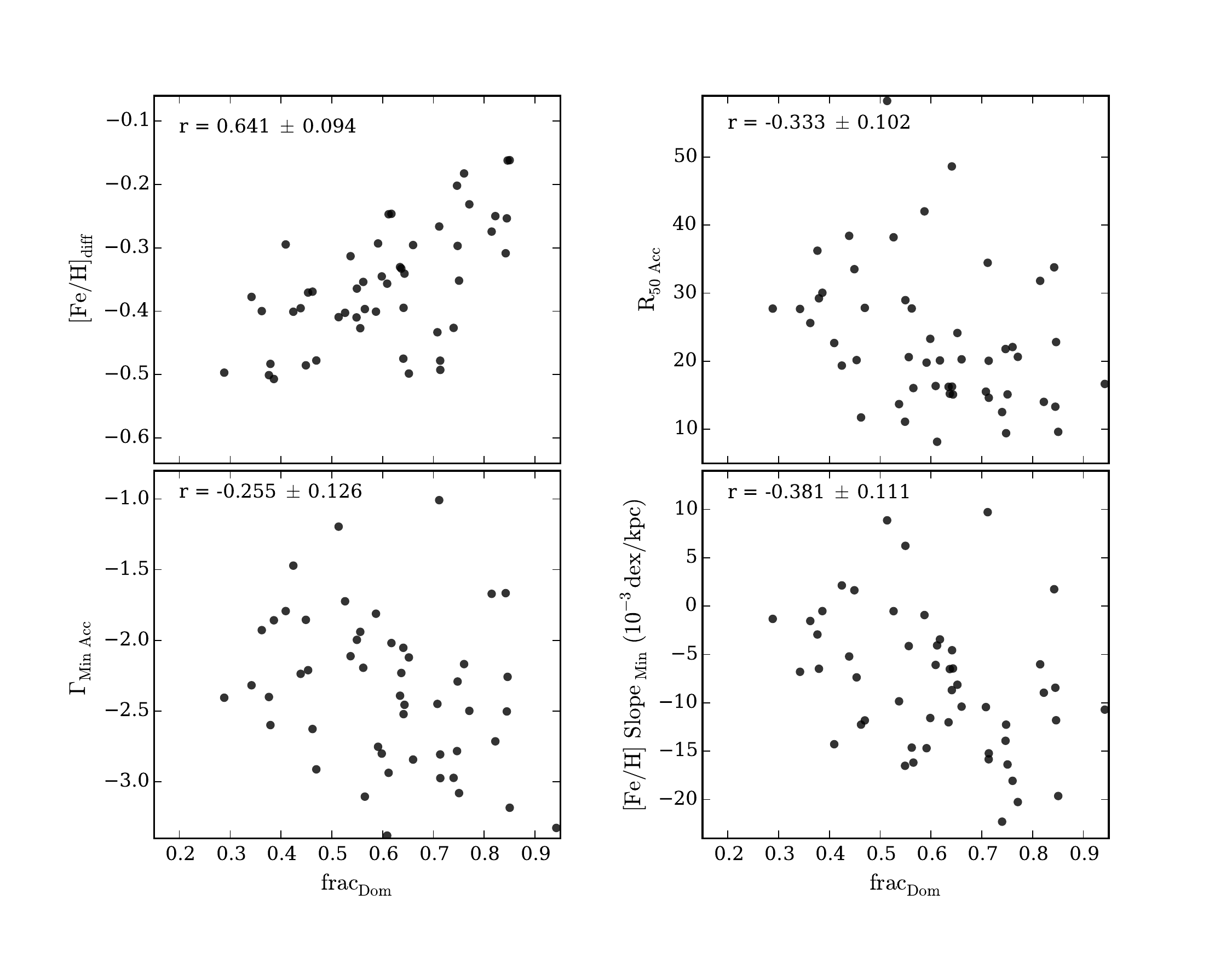}
\caption{For MW-like mass galaxies with an accreted stellar mass in the range 
$9.5 < \log\, \mathrm{M_{acc}}  < 9.7$, we plot the correlations between 
$\mathrm{frac_{Dom}}$ and $\mathrm{[Fe/H]_{diff}}$, $\mathrm{R_{50\,acc}}$, 
$\mathrm{\Gamma_{acc\,min}}$ as well as the metallicity gradient of the accreted
component along the minor axis. We also calculate the Pearson Correlation 
coefficient (r) for the relationships. The metallicity of the accreted stellar component
is highly informative about $\mathrm{frac_{Dom}}$, while its physical size, its 
surface mass density slope as well as its metallicity gradient contains little or no 
information.} 
\label{fig_domfrac_predict}
\end{figure*}

On the other hand, spatial information encoded in the physical properties
of the accreted stellar component contains little information about 
$\mathrm{frac_{Dom}}$.  In Figure \ref{fig_domfrac_predict}, we demonstrate 
that for a narrow accreted  stellar mass range ($9.5 < \log\, \mathrm{M_{acc}}  
< 9.7$ for MW-like mass galaxies), $\mathrm{frac_{Dom}}$ correlates most with  
$\mathrm{[Fe/H]_{diff}}$ in comparison to other indicators involving spatial
information.  This confirms that the spatial physical properties of the accreted
stellar halo contained in the power-law density slope and metallicity
gradient cannot constraint $\mathrm{frac_{Dom}}$ alone.

\subsection{Summary}
In general, we have demonstrated that it is possible to constrain the 
characteristics of the dominant progenitor in terms of its stellar mass 
as well as the time of its accretion. The accreted stellar metallicity
can best constrain the mass of the dominant progenitor, while the 
physical spatial information of the accreted stellar component can best constrain
the time of accretion of the dominant progenitor.

\section{Observable measurements along the minor axis}
\label{sec:minor}
We have seen how the metallicity and the mass of the total accreted stellar component 
can inform us about the characteristics of the dominant progenitor. Observationally, however, 
we have no direct access to either of these two quantities. Rather, we rely on 
observational proxies for the same: accreted stellar metallicity estimated at a fixed 
distance along the minor axis and aperture measurement of the mass of the accreted stellar 
component. In this section, we explore how we can recover the metallicity and mass of the 
total accreted stellar component from the observational proxies. Although we concentrate 
on MW-like mass galaxies for illustrative purposes, the results of this 
section are generalizable to galaxies spanning a broad range in masses. 

\subsection{Accreted Stellar Metallicity}
MW-like mass galaxies exhibit a range of stellar metallicity gradients along the 
minor axis even at large galacto-centric distances \citep{Monachesi2016a}. A similar trend
is found in Illustris galaxies: they show a similar range in metallicity gradients 
along  the minor axis even in the accreted component (see Figure \ref{fig_MW_diversity}).
This suggests that a measurement of the metallicity of the accreted component along
the minor axis at large galacto-centric distances can be considerably lower than the total 
accreted stellar metallicity for galaxies with high $\mathrm{frac_{Dom}}$. The 
presence of a metallicity gradient informs us that a measurement along the minor axis 
is not representative of the metallicity of the total accreted stellar component, as well as about 
the degree of that misrepresentation. The information encoded in the metallicity 
gradient could allow us to reconstruct the metallicity of the total accreted stellar component 
from the metallicity measurement along the minor axis.

In order to directly compare our results with the estimated metallicities 
from the  GHOSTS-like observations, we consider the median accreted stellar 
metallicity  along a projected wedge of 30 degrees along the minor axis 
between a  galacto-centric distance of 25 and 45 kpc. 

\begin{figure}
\centering
\includegraphics[width=0.5 \textwidth]{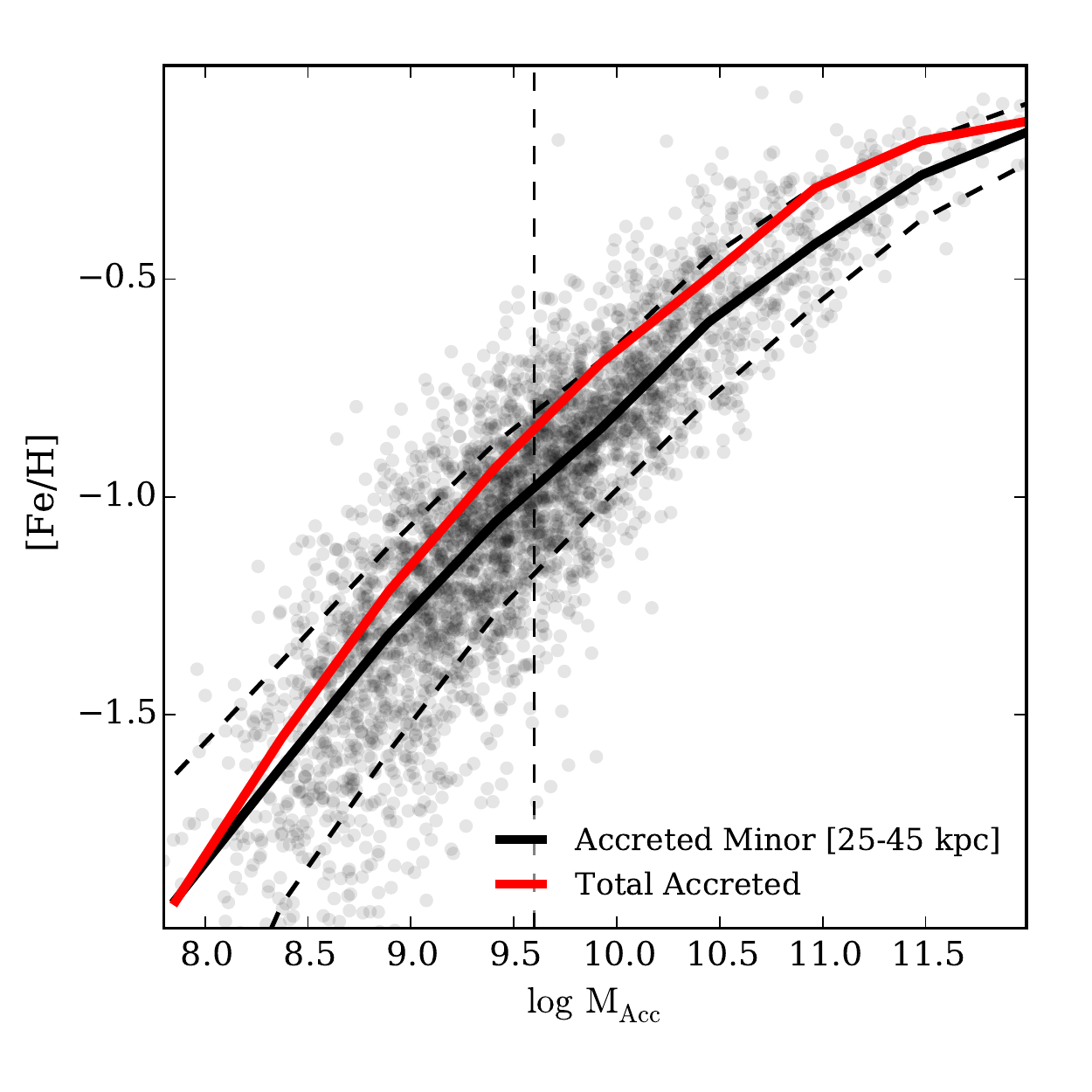}
\caption{The median metallicity of the accreted component (estimated in wedges 
with 30 degrees along the minor axes between a galactocentric distance of 25 and 45
 kpc) as a function of the accreted stellar mass of the Illustris galaxies. The black solid 
 and dashed lines indicate the median and the 16/84th percentile of the distribution. The red solid line indicates the median of the metallicity of the total accreted stellar component. The metallicity estimated along the minor axis will be lower than the metallicity of the total accreted stellar component.} 
\label{fig18}
\end{figure}

Before we examine MW-like mass galaxies, we first plot the median metallicity 
of the accreted stellar particles along a projected wedge along the minor axes 
between 25 and 45 kpc, as a function of accreted stellar mass for all central
Illustris galaxies in Figure \ref{fig18}. This relationship is similar to the 
total accreted metallicity-stellar mass relationship (compare to Figure \ref{fig3}): 
While both relationships have similar slope, the relationship along the 
minor x-axis is $\sim$0.2\,dex lower. At lower accreted 
stellar mass ($<\mathrm{\log\,M_{acc}}$), the relationship along the 
minor axis approaches the  total accreted metallicity-stellar mass relationship. 
There is a considerable spread in the relationship along the minor 
axis, comparable to that found along the total accreted metallicity-stellar mass
relationship in Figure \ref{fig3}. The difference in the two relationships can be 
accounted for in terms of metallicity gradients along the minor axis.

\begin{figure}
\centering
\includegraphics[width=0.5 \textwidth]{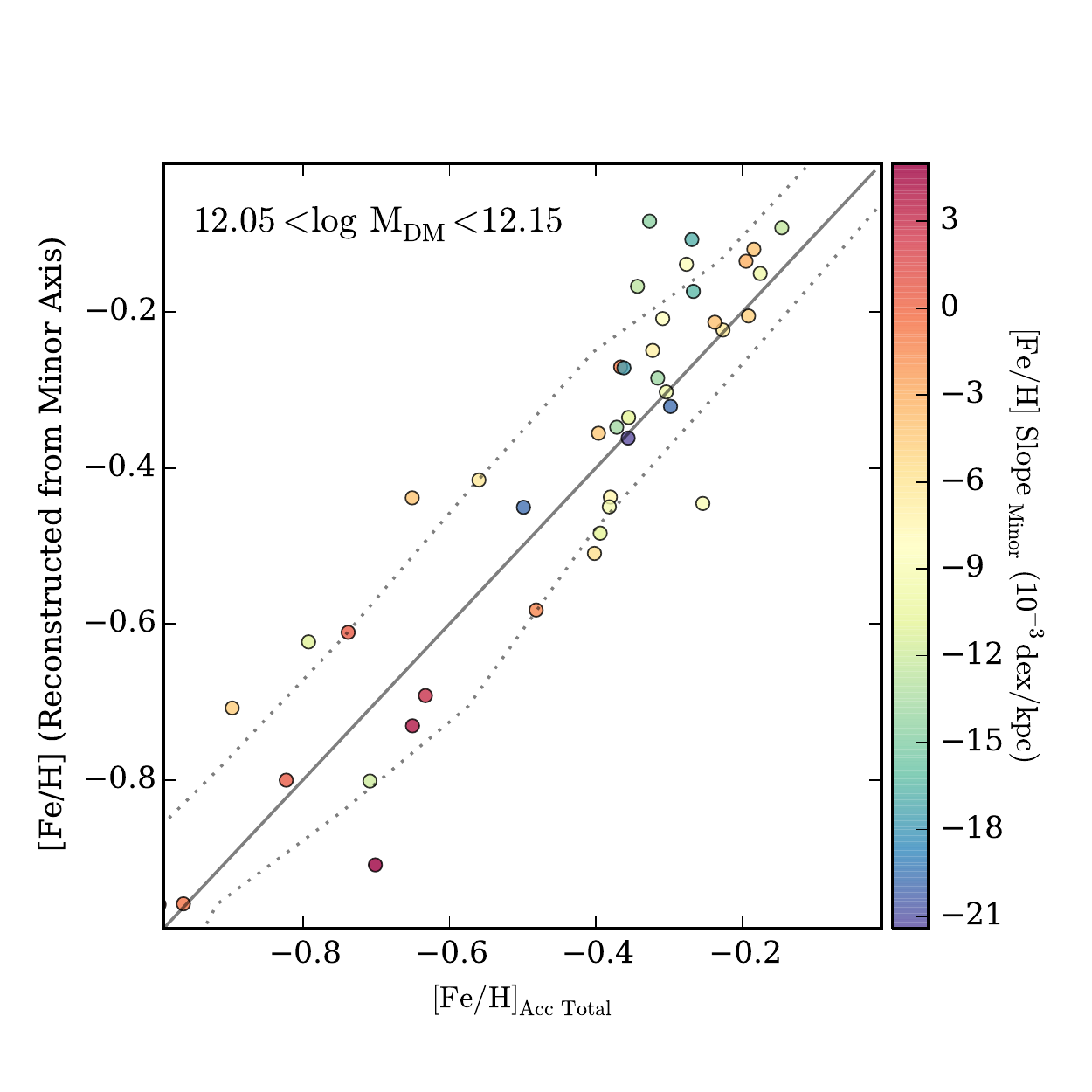}
\caption{The reconstructed metallicity of the total accreted stellar component as 
a function of the true median accreted stellar metallicity colour-coded by the 
metallicity gradient for MW-like mass galaxies. The accreted metallicities are 
reconstructed using a simple linear extrapolation using the estimated accreted 
stellar metallicity along the minor axis at 35 kpc as well as the information 
encoded in the metallicity gradients. The dashed lines shows the scatter in 
the relationship. The metallicity of the total accreted stellar component can be 
reconstructed from information about the metallicity along the minor axis.}
\label{fig19}
\end{figure}

We now proceed to examine if we can reconstruct the total accreted
stellar metallicity from observations along the minor axis using the 
metallicity gradients for MW-like mass galaxies. As a first order, we 
consider a simple linear extrapolation  scheme such that $\mathrm{[Fe/H]_{reconst}}\,=\,
\mathrm{[Fe/H]_{minor}}\, +\,\mathrm{gradient} \times \mathrm{35\,kpc}$. 
In Figure \ref{fig19}, we explore the ability of this simple linear 
extrapolation scheme to recover the metallicity of the total accreted stellar component. 
While there is a scatter of the reconstructed accreted metallicity around 
the total accreted metallicity, this scatter is around 0.1 dex. This lends 
support to the idea that  we can reconstruct the metallicity of the total accreted stellar 
component within an accuracy of 0.1 dex from the information encoded 
in the metallicity  gradients. More sophisticated extrapolation schemes 
using the information encoded in the density profiles could do better.

\subsection{Accreted Stellar Mass}
\begin{figure*}
\centering
\includegraphics[width=0.325 \textwidth]{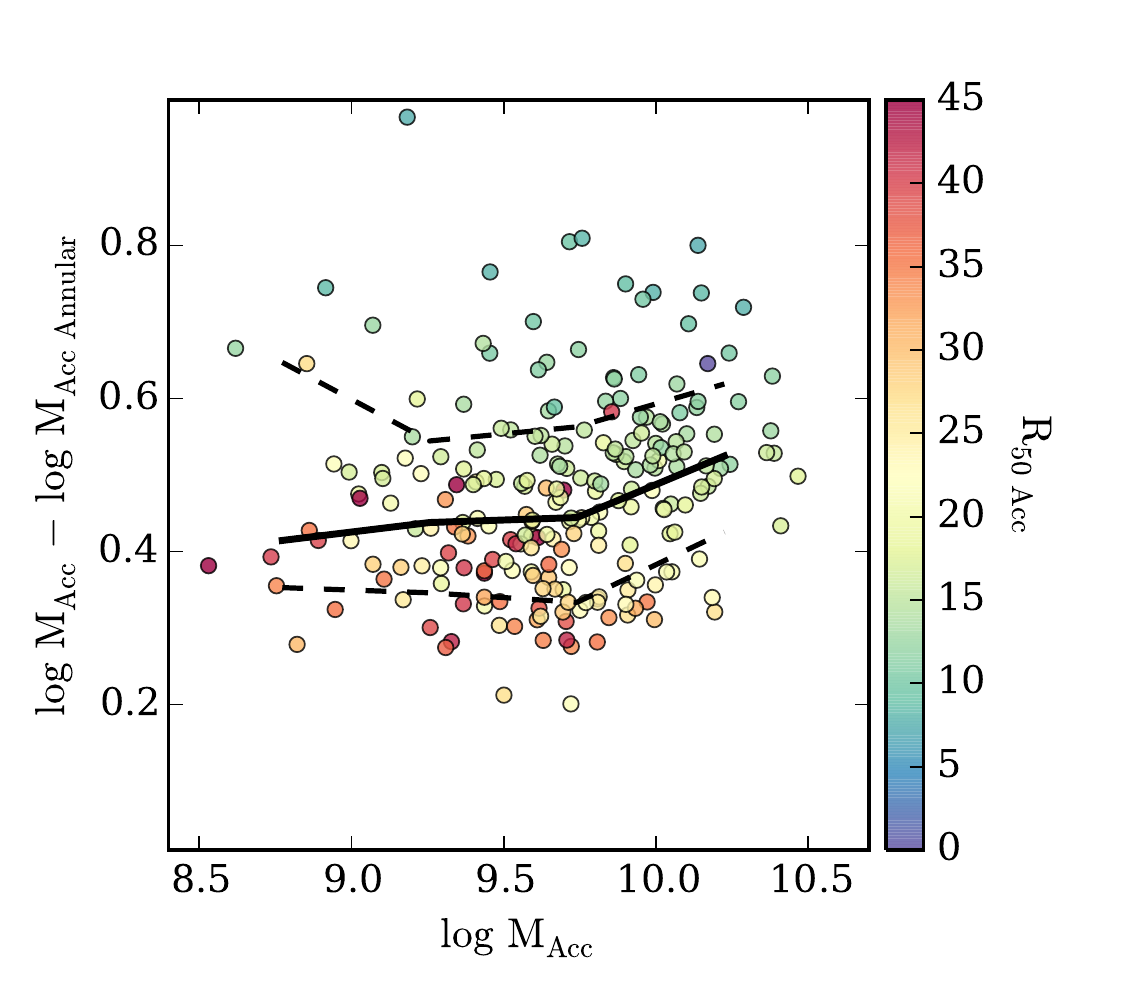}
\includegraphics[width=0.325 \textwidth]{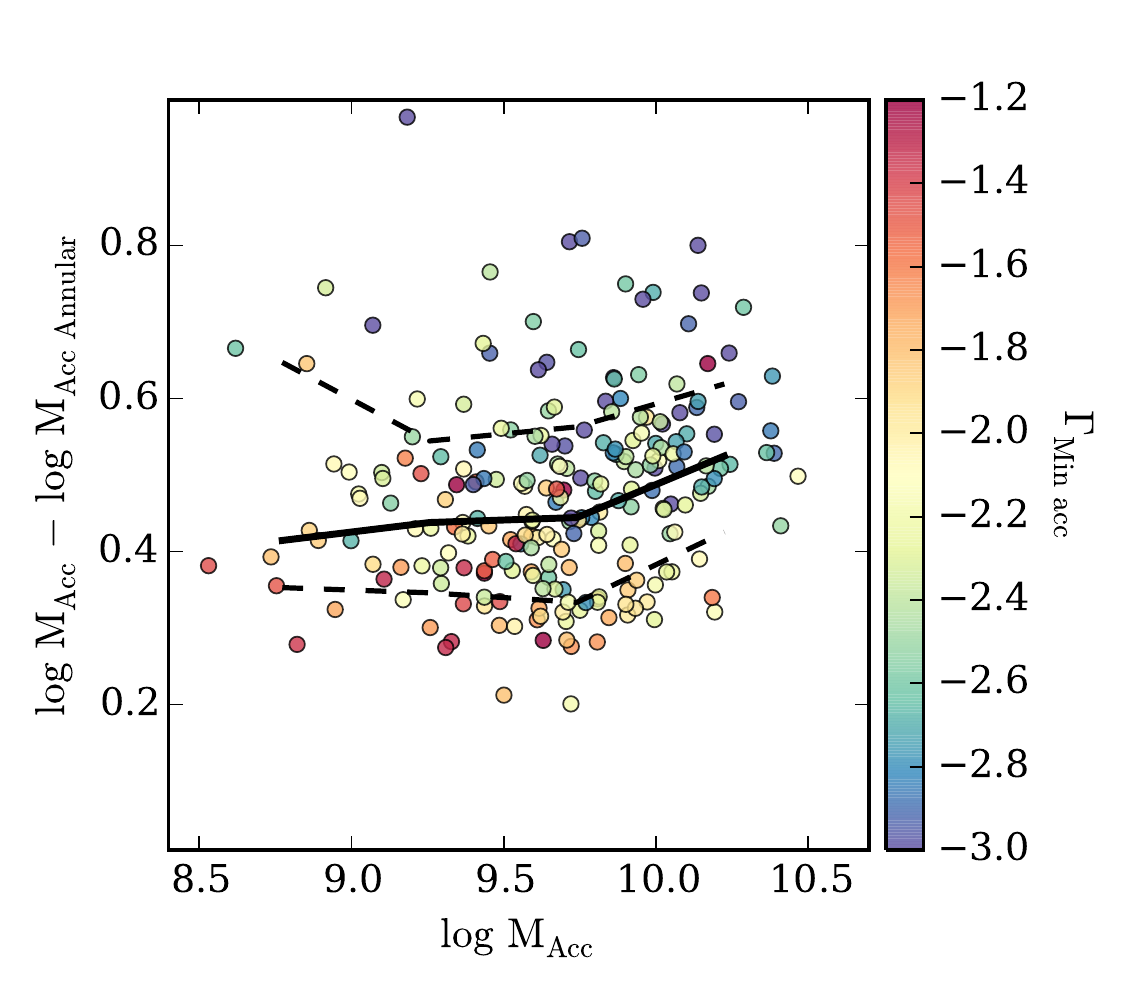}
\includegraphics[width=0.325 \textwidth]{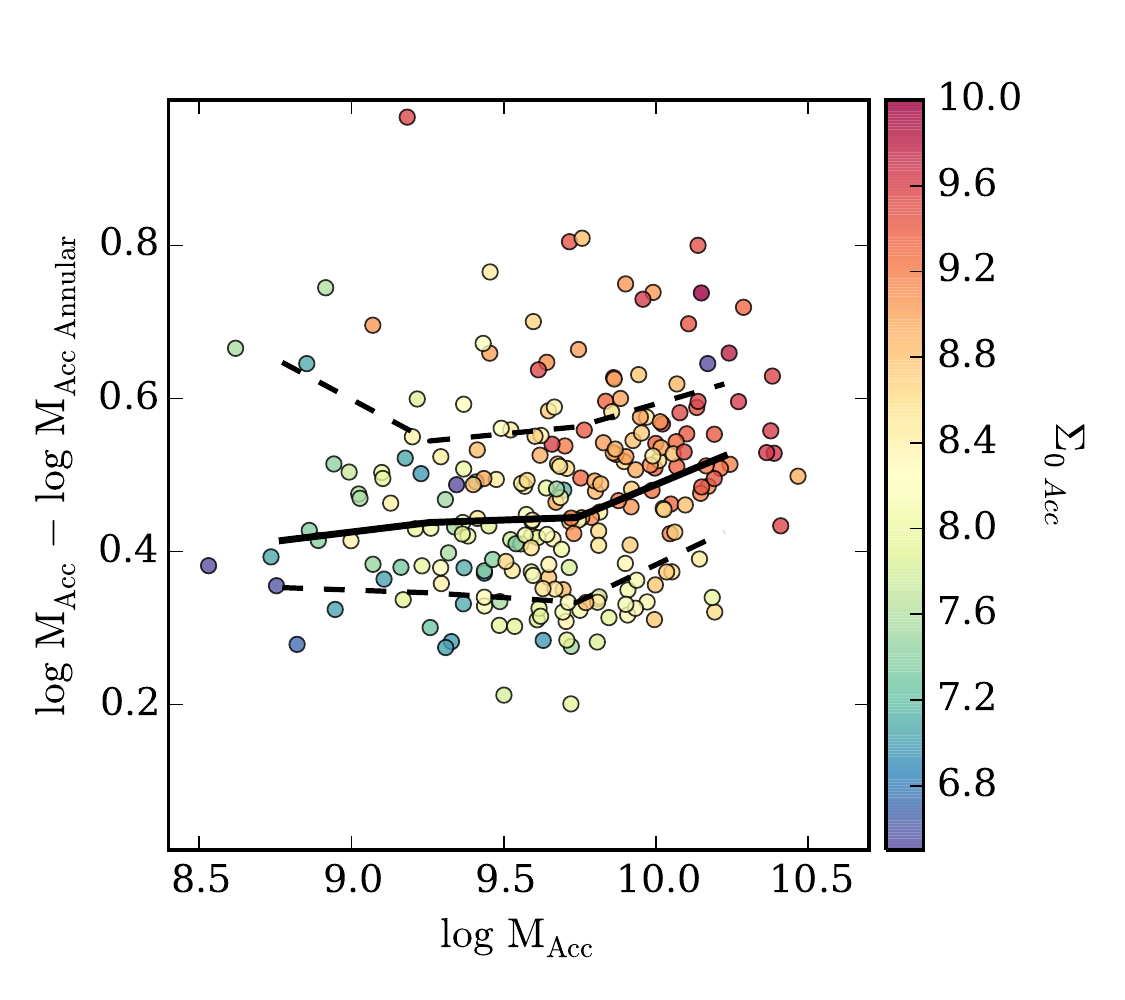}
\caption{The difference between the total accreted stellar mass and the "aperture" 
measurement of the accreted stellar mass (between  10 and 50 kpc) as a function 
of the total accreted  stellar mass for MW-like mass galaxies. The points are 
colour-coded according to 3D half-mass radius $\mathrm{R_{50\,Acc}}$ (left), 
the 2d power-law density slope of the accreted stellar component along the 
minor axis $\mathrm{\Gamma_{Min\,acc} }$ (center), and the normalisation of 
the power-law fit of the density profile $\mathrm{\Sigma_{0\,acc}}$ (right). 
The total accreted stellar mass can be reconstructed from aperture mass 
measurements of the stellar halo to within an accuracy of 0.15 dex. }
\label{fig20}
\end{figure*}

Because much of the accreted stellar material of a galaxy is not directly observable, 
measuring its total accreted stellar mass involves model-dependent assumptions. 
Using accretion-only models of \cite{Bullock2005}, \cite{Harmsen2017} and 
\cite{Bell2017} extrapolated an ``aperture" stellar halo mass between 10-40\,kpc
along the minor axis to a total accreted stellar mass. We evaluate these
extrapolations for Illustris MW-like mass galaxies.  To do this, we consider a  
circular aperture measurement (between 15 and 50\,kpc along the minor axis, 
corresponding approximately to an elliptical aperture of semi-minor axis 10--40\,kpc) of 
the stellar mass of the accreted halo. In Figure \ref{fig20}, we plot the 
difference between the total accreted stellar mass and the aperture measurement 
of the same as a function of the total accreted stellar mass. We find
that the total accreted stellar component mass is $\sim 3\,\times$ the ``aperture"
accreted mass with a scatter of 30\% and a $<15\%$ variation with accreted stellar
mass.  Consequently, the total accreted stellar mass can be reconstructed to within an
accuracy of 0.15 dex. This validates the claims 
of \cite{Harmsen2017} and \cite{Bell2017}, and allows us to connect the observable 
part of the accreted stellar component at large galacto-centric distances with the total 
accreted stellar material, the bulk of which is present at small radii due to  
dynamical friction \citep{Cooper2013,Amorisco2017}.

Moreover, we find that the scatter in the difference between the total and
the ``aperture" measurements at a fixed accreted stellar mass correlates best
with  the 3D half-mass radius of the accreted stellar component ($\mathrm{R_{50\,Acc}}$). 
Galaxies with a more extended stellar halo have a smaller aperture accreted 
stellar mass compared to galaxies with a less extended stellar halo. 
Similar correlations are also seen with the 2d power-law density slope of the 
accreted stellar component along the minor axis ($\mathrm{\Gamma_{Min\,acc} }$)
as well as its normalization $\mathrm{\Sigma_{0\,Acc}}$ (see Figure \ref{fig20}).
As demonstrated in Section \ref{sec:diversity}, $\mathrm{R_{50\,Acc}}$ at a fixed 
accreted stellar mass correlates best with the time of accretion of the dominant merger.

Thus, assuming that the stellar halo structures in Illustris are reasonably accurate, 
we suggest that future analyses using information encoded in the minor axis density 
profile of the accreted stellar component may permit more robust extrapolations from 
an ``aperture" measurement to the total accreted stellar mass of the galaxy.

\section{Discussion}
\label{sec:discuss}

In this work, we use the Illustris simulations to build intuition of how
the accreted metallicity-stellar mass relationship comes about in central galaxies and 
how it can inform us about the properties of the dominant progenitor. 
A powerful advantage of Illustris over previous works \citep[e.g.,][]{Renda2005,Font2006,Deason2016} 
is the large dynamic range in DM halo mass, and its use of a large cosmological volume.  
This allows us to explore galaxies over a broad range in accretion histories.
We find the accreted metallicity-stellar mass relationship extends over several
orders of magnitude in accreted stellar mass. The large dynamic range of this 
relationship is truly remarkable. Due to the galaxy metallicity-stellar mass 
relationship, the dominant progenitor drives the metallicity of the accreted 
stellar halo \citep{Deason2016}. This relationship enables us to probe the 
complex multi-dimensional space of an accretion history of the galaxy in terms 
of its dominant progenitor.

We use Illustris to discuss possible observational metrics of the properties 
of the dominant progenitor of a given stellar halo. The metallicity of a stellar 
halo is driven by its dominant progenitor, and together with the total accreted 
mass gives an estimate of the fraction of the total stellar halo contributed by 
the most dominant progenitor. Metallicity and density gradients allow one to 
constrain the time of the dominant accretion; galaxies with earlier dominant 
accretions have more compact halos with steeper gradients, when compared to 
the ensemble of stellar halos with a given accreted mass. We use Illustris to 
argue that minor axis observations of resolved stars in stellar halos contain 
enough information to recover total accreted mass to better than 0.15\,dex 
accuracy, and median accreted metallicity to within 0.1\,dex, opening the 
door to applying these metrics to observational datasets.

\subsection{Limitations of this study}
The strengths of Illustris for this work are its combination of volume, resolution, 
dynamic range in halo masses, a reasonable satellite progenitor population and 
tagging of star particles as accreted or {\it in situ}. Yet, all simulations 
have limitations, and Illustris is no exception. Illustris has four potentially 
important limitations that impact our results. 
First, the stellar mass function (SMF) of the Illustris simulations shows an 
excess of low stellar mass galaxies up to redshift $z \sim 2$. Secondly, the 
stellar mass-halo mass relationship of the Illustris galaxies match the 
observational constraints imperfectly. Thirdly, the  Illustris simulation has 
a steeper-than-observed galaxy metallicity-stellar mass relationship, 
exhibiting little time evolution. Finally, the Illustris galaxies fail
to reproduce the observable mass-size relationship of galaxies, galaxies
being fairly large for a given stellar mass \citep{Snyder2015}.

Despite these limitations, the main results of this paper are fairly robust,
since we focus only on the accreted stellar component. As can be seen by 
comparison with e.g., \citet{Deason2016} or \citet{Amorisco2017}, the main 
requirements for this analysis are a reliable dark matter framework, a 
reasonable stellar mass for halos at a given dark matter mass and a reasonable 
metallicity. Illustris satisfies these requirements easily. 
For MW-like mass galaxies,  the bulk of the accreted stellar material is 
contributed by large progenitors \citep[see also][]{Deason2016}, with DM halo  
masses $\sim 4 \times 10^{11} M_{\odot}$ \citep{Purcell2007}. At this mass, Illustris 
reproduces the stellar-to-halo mass ratio reasonably well \citep[][]{Vogelsberger2014a}.
Smaller subhaloes deviate somewhat from the stellar mass--halo mass relation, 
but do not contribute sufficiently to the final stellar halos to adversely bias the 
accreted stellar masses of central galaxies \citep[see Fig. 1 of ][]{Purcell2007}. 
The Illustris simulation has a strong metallicity--mass relation. Once we adjust 
the normalization of the metallicity--mass relation  to take into account the 
[M/H] to [Fe/H] conversion and any deficiencies in the Illustris chemical 
evolution framework (Fig.\ \ref{fig1}), it reproduces the metallicity of galaxies 
to within $\sim 0.1$\,dex, and the resultant stellar halos are an 
acceptable match to observations (Fig.\ \ref{fig18b}).  Moreover, 
the spatial distribution of the accreted stellar material is governed
solely by gravity and dynamics. These suggest that the {\it qualitative} 
(and to a certain extent, the quantitative) results of our analysis are robust. 
In order to develop further our ability to quantitatively mine stellar halos 
for the properties of the dominant progenitor, it will be important to map 
out the empirical ranges of metallicity, metallicity gradients, masses and 
density gradients as a function of virial mass for a sizeable sample of galaxies. 
Relative trends will be more robust than absolute numbers, and these can be 
revealed only in a diverse sample which exercises the dynamic range of 
accreted halos. These insights may require some adjustments in the analysis of 
observations. For example, if the galaxy metallicity-stellar mass relationship evolves with 
time, $\mathrm{[Fe/H]_{diff}}$ will be a function of both the mass of the 
dominant progenitor and the time of its accretion. In such a case, a joint 
analysis of the $\mathrm{[Fe/H]_{diff}}$  with the spatial information of the 
accreted stellar component would be necessary to put constraints of the mass of the 
dominant progenitor and the time of its accretion. 

When comparing with observations, the presence of an {\it in situ} halo 
\citep{Zolotov2009,Cooper2015} could complicate the measurement of the 
accreted stellar metallicity. This motivates our reliance on comparisons 
with minor axis measurements of disk-dominated galaxies from GHOSTS 
\citep{Monachesi2016a,Harmsen2017}, where contributions from {\it in situ} 
stars are expected to be negligible \citep{Pillepich2015,Monachesi2016b}. 
There is a substantial debate among simulators today as to the amount 
of {\it in situ} stars at large radius in simulated galaxies. The fraction 
of {\it in situ} stars present at large galacto-centric distances is highly 
dependent on the nature of feedback present in the hydro-dynamical simulations.  
The Illustris galaxies have particularly extended stellar haloes due to the 
presence of a large fraction of {\it in situ} stars at large galacto-centric 
distances \citep{Pillepich2014}. Such prominent {\it in situ} halos would 
over-produce the observed 10--40\,kpc stellar masses of real galaxies 
(Fig.\ \ref{fig18b}; see also \citealp{Harmsen2017}). Furthermore, the 
flat observed metallicity gradients found in half of the GHOSTS galaxies 
is inconsistent with ubiquitous extended {\it in situ} halos in MW-like 
mass galaxies \citep{Monachesi2016a,Harmsen2017}. Finally, the presence 
of a prominent {\it in situ} halo would reduce the slope of accreted 
metallicity-stellar mass relationship for MW-like mass galaxies towards 
the slope of the  galaxy-metallicity stellar mass relationship, in seeming 
conflict with the observed relationship \citep[see also Fig.\ \protect\ref{fig18b}]{Harmsen2017}.

Finally, because this work was motivated by the template of  GHOSTS-like observations 
with HST, the main results of this study have restricted applicability to nearby edge-on 
galaxies. Realistically, the number of inclined near-by edge-on galaxies we can observe with 
HST are limited. New space missions including the  Wide Field Infrared Survey Telescope (WFIRST) 
as well as  the  James Webb Space Telescope (JWST) will help to increase the sample size beyond 
the current limitations of 10 Mpc. Given the  limited number of edge-on systems within this 
volume, it is crucial we understand how to extend these techniques to deal with face-on 
galaxies. Improving our modelling schemes in dealing with an {\it in situ} stellar component
will help us disentangle the {\it in situ} and accreted stellar components of nearby galaxies.

\subsection{Implications and comparison with other works}

While previous works showed a relationship between the mass of an accreted halo and 
its metallicity \citep{Renda2005,Font2006,Deason2016}, Illustris allowed us to characterize 
the accreted metallicity-stellar mass relationship over a large range in DM halo mass, accretion 
histories and in a cosmological volume. In accord with previous works, the main driver of this 
relationship is the {\it galaxy} metallicity-stellar mass relationship, where the dominant 
progenitor drives both the metallicity and the mass of the accreted stellar component \citep{Deason2016}. 

The scale and dynamic range of Illustris allow us to explore the scatter in the accreted
metallicity-stellar mass relationship, finding that Illustris predicts this scatter to be 
both considerable and informative. This finding was hinted at by \citet{Deason2016}, but 
is placed on a substantially firmer qualitative and quantitative footing by Illustris. 
The accreted metallicity gives an informative measure of $\mathrm{M_{Dom}}$; this in 
practice can correspond to a wide range in $\mathrm{M_{acc}}$ depending on the merger 
history (Fig.\ \ref{fig3}). This is of possible importance for those wishing to 
characterize the total accreted mass using metallicity only; such inferences will 
be very uncertain without the addition of extra information (e.g., a halo mass 
estimate, following Fig.\ \ref{fig3}). 

 In this work, we chose to parametrize the accretion history of a galaxy 
in terms of the mass and time of accretion of the most dominant progenitor. 
In particular, we concentrate on ($\mathrm{frac_{Dom}}$), the fraction of accreted 
stellar material contributed by the most dominant progenitor. There are many 
equivalent ways of parametrizing the accretion history of the galaxy, for example,
in terms of the mean merger ratio \citep{Rodriguez-Gomez2015,Cook2016}, the mass 
weighted average destroyed satellite mass \citep{Deason2016} or the number of 
significant progenitors which contribute a given percentage of the mass of the stellar halo 
\citep{Cooper2010}. Our choice of concentrating on the dominant progenitor is motivated 
by the fact that it drives the metallicity of the total accreted stellar 
component.

Previous works have argued that the properties of the largest progenitor can be 
constrained by using the accreted stellar mass \citep{Deason2016,Amorisco2017b}, 
the slope of the density profile \citep{Pillepich2014}, or metallicity gradients 
\citep{Cooper2010,Tissera2013,Hirschmann2015}. Our analysis of Illustris suggests 
that the stellar halo metallicity is a much more direct probe of the mass of the 
largest progenitor, and that the scatter in accreted stellar metallicity at a given 
accreted stellar mass directly reflects the fraction of the accreted stellar mass 
($\mathrm{frac_{Dom}}$) contributed by the most dominant progenitor. We argue that 
while the alternative metrics suggested by other authors {\it do} correlate with 
dominant progenitor mass, such correlations are contributed to by confounding quantities.
At a given accreted stellar mass, one can have a range in $\mathrm{frac_{Dom}}$ (see Figure 
\ref{fig8}). Both the density profiles and the metallicity gradients are dependent 
not only on the mass of the progenitor but also on the concentration of its DM halo and
the time of its accretion. On the other hand, the metallicity of the total accreted stellar 
component depends on the mass of the progenitor alone. We note that this result 
is somewhat sensitive to time evolution of the galaxy stellar metallicity--stellar mass 
relation; if this relation evolves, a joint analysis of the stellar halo metallicity 
in conjunction with its gradients will be necessary to jointly constrain the mass 
and time of accretion of the dominant progenitor. 

For galaxies where it is  difficult to estimate its accreted stellar mass
(e.g. early-types or face-on disk galaxies), a measurement of the total accreted 
stellar metallicity can still constrain the mass of the dominant progenitor, for 
galaxies in a fixed stellar mass range (see Figure \ref{fig21}). If there is an
evolution in the galaxy metallicity-stellar mass relationship with time, upper 
limits can be set on the mass of the dominant progenitor.

Recently, \cite{Amorisco2017b} showed that accreted stellar mass in addition to
the number of distinct surviving massive satellites can also provide constraints
on the accretion history of the galaxy. \cite{Deason2016} also explore similar 
ideas. This requires surveying around the galaxy for its  massive satellites
in addition to measuring its accreted stellar component. This may  be viable 
for nearby galaxies using wide-field observations of diffuse light 
\citep{Merritt2016} or resolved stars \citep{Carlin2016}.

Dynamical friction affects the radial mass distribution  of tidally stripped 
material \citep{Amorisco2017}. Large accreted satellites deposit their material 
at the  centre of the galaxy, while smaller progenitors are preferentially 
deposited at large galacto-centric distances, affecting the density profiles
of accreted stars. This leads to a preferential metallicity segregation, 
visible in metallicity gradients in the accreted stellar component.

We find that quantities encoding spatial information of the accreted stellar
halo including  the power-law density profile and metallicity gradient of the 
accreted stellar component becomes steeper with increasing mass of the dominant 
progenitor (see Figure \ref{fig_MW_diversity}). This is consistent with 
the analysis of \cite{Amorisco2017}.  Although the spatial information (density 
slopes and metallicity gradients) is  primarily driven  by $\mathrm{frac_{Dom}}$ 
(or increasing accreted stellar mass), the large scatter in these relations do 
not allow  us to predict the mass of the  most dominant satellite using purely spatial 
information alone. Rather, we find that at a given accreted stellar mass, the scatter 
in the spatial information correlates best with the time of accretion of the dominant 
progenitor. Galaxies which accrete earlier in time tend to have a more compact and 
steeper accreted stellar components than those that accrete later in time, at a fixed 
accreted stellar  mass. These results are also consistent with the particle-tagging 
simulations of  \citet{Cooper2010} which showed that their metallicity gradients 
of the accreted component correlate with the mass of the dominant progenitor. 

The fundamental drivers of metallicity gradients in the accreted stellar component are
the intrinsic metallicity gradients in the massive accreted satellites. The massive 
accreted satellites in the Illustris simulations have metallicity gradients leading 
to gradients in the accreted stellar haloes of galaxies. Similarly, in the 
simulations of \citet{Cooper2010}, due to the particle-tagging methodology employed 
there, the massive accreted satellites had metallicity gradients. On the other hand, 
the particle-tagging simulations  of \cite{Bullock2005} showed hardly any metallicity 
gradients in the accreted stellar component \citep{Font2006}. This can be traced back 
to the limitations of the tagging procedure used in these simulations: the accreted 
satellites did not possess any intrinsic metallicity gradients.

It would be interesting to compare the metallicity gradients of the accreted stellar 
component of the Illustris MW-like galaxies with the observational constraints. 
Since \cite{Monachesi2016a} did not publish any metallicity gradients for the GHOSTS galaxies, 
we cannot undertake a quantitative comparison. However, their published colour gradients 
allows us a qualitative comparison. Their colour gradients span from 0 to -0.004 
mag/kpc \citep{Gilbert2014}. M31 is one of those galaxies with an extremely large 
metallicity gradient. This suggests that the gradients in the GHOSTS galaxies span 
from 0 to -0.01 $\mathrm{dex\,kpc^{-1}}$. From the right panel of Figure \ref{fig_MW_diversity}, 
we find that the metallicity gradients of the accreted stellar component of MW-like 
Illustris galaxies spans 0.005 to -0.015 $\mathrm{dex\,kpc^{-1}}$. This is consistent 
with the metallicity gradients of the GHOSTS galaxies.

\subsection{How do these results apply to systems with a substantial {\it in situ} stellar halo?}
In order to relate these results to a broader set of observations, one needs to 
include the modelling of {\it in situ} stellar material.  Because the results 
presented in this work are only valid for accretion--only systems,  our best bet
for observing these trends are  with GHOSTS-like HST observations of the minor axis
of nearby edge-on galaxies.  However, some of the results of this work 
may also be valid for other galaxies in the limit of large galacto-centric distances,
which are dominated by accreted-only stars.

Early cosmological hydrodynamical simulations had a hard time reproducing realistic
galaxies, and correspondingly, the {\it in situ} component. Nevertheless, much 
progress has been made in recent years. \cite{Font2011} showed the observable 
metallicity gradients in the inner part of the galaxies ($<\,30\,\mathrm{kpc}$) 
are related to the relative fraction of {\it in situ} to accreted stellar material 
(or the accreted stellar fraction). \cite{Cooper2013} using particle-tagging 
simulations extended this analysis and  showed for a range in stellar masses that 
the shape of the density profile encodes information about  the fraction of accreted 
stellar material. \cite{Pillepich2014} obtained qualitatively similar results 
using the Illustris hydrodynamical simulations by parametrizing the density profile 
with a power-law fit. Galaxies with a higher accreted stellar fraction have a shallower 
slope.

However, the question of the ability of ``observable'' density profiles and metallicity 
gradients in galaxies (involving contamination of {\it in situ}  stellar material)
to differentiate between possible accretion histories of a galaxy (in terms of merger 
fraction or time of accretion) is an open one and subject to much debate, especially
in the limit of large galacto-centric distances. While a number of hydrodynamical 
simulations show that the density profiles and the metallicity gradients of the accreted 
stellar component  encodes information  of the accretion history of the galaxy 
\citep{Font2011,Tissera2013,Hirschmann2015}, they differ in the predictions of 
their {\it in situ} stellar component and  its radial extent, and how much it 
dominates the overall profiles. 

In simulations which over-predict the amount of the {\it in situ} stellar halo,
analyses of their radial profiles at smaller radial scales ( $<\,30\,\mathrm{kpc}$)
will be biased towards washing out any signal of the accretion history of the galaxy.
Indeed, \cite{Cook2016} find that the metallicity gradients and the density slopes 
(between 2-4 $R_{1/2}$) of quiescent Illustris  galaxies are only sensitive to 
the  accreted stellar fraction. 

On the other hand, analyses which extend to larger radii where the accreted 
stellar material is predominant  find contrasting results. \cite{Hirschmann2015} find 
that the  metallicity gradients of quiescent galaxies (between 2-6 $R_{1/2}$) are 
correlated with the merger-mass  ratio \citep{Hirschmann2013}. It must also be noted 
that the simulations of \cite{Hirschmann2013} do a better job in reproducing 
the mass-size relationship of galaxies in contrast to the  Illustris 
simulations \citep{Snyder2015}. Recent simulations of  disk galaxies also 
indicate that the  observable outer metallicity gradients are indeed correlated 
with the accretion history of the galaxy \citep{Tissera2013}. 

We conclude that at large radii (between 4-6 $R_{1/2}$) where galaxies are dominated 
by their accreted stellar material, the observable density profiles and the metallicity 
gradients at these distances continue to retain information about the accretion history of 
a galaxy. At intermediate distances (between 2-4 $R_{1/2}$), the accretion information 
content of the density profiles and metallicity gradients of galaxies depends upon the 
amount of {\it in situ} stellar halo being produced in the simulations. 

\subsection{Reflections on the Milky Way, M31 and NGC 5128}
While future works will explore the implications of our results on the 
stellar halo measurements for individual nearby galaxy merger histories 
in considerably more depth (see, e.g., \citealp{Bell2017} for a first 
attempt), it is interesting to consider some implications for our 
understanding of the Milky Way, M31 and NGC 5128. We are on firmer 
ground making inferences about the Milky Way and M31, as GHOSTS provides an 
observed sample of peer disk galaxies, allowing relative statements to be made. 
In particular, the Milky Way's and M31's properties are relatively extreme, lying 
at the low and high-mass ends of the distribution (e.g., Fig.\ \ref{fig18b}). 
For NGC 5128, no such sample of peers exists, limiting our ability to make 
strong inferences about accretion history. 

\subsubsection*{The Milky Way} 

As many measurements of the MW stellar halo's properties do not incorporate 
Sagittarius, or account for it only incompletely, we discuss the Milky Way 
neglecting Sagittarius first, then reflect on its possible importance at the end. 

The MW distinguishes itself amongst nearby galaxies with similar total stellar 
masses by having a very small stellar halo, consistent with a low accreted stellar
mass. The total accreted stellar mass of the MW has been estimated to 
be $\mathrm{M_{Acc}} \sim 4-7 \times 10^{8} \mathrm{M_{\odot}}$ (\citealp{Bland-Hawthorn2016}, 
following \citealp{Bell2008}). Such a  low accreted stellar mass is consistent 
with the idea that our Galaxy has been largely undisturbed over the past several 
gigayears and  has accreted very little compared to similar sized galaxies.  The 
accretion history of the Galaxy is more an exception than the norm for 
MW-like mass galaxies, lying at the low tail of the distribution of accreted 
stellar mass (see Figure \ref{fig8}). We do not expect to find many galaxies
with similar sized stellar haloes, but there may still be some analogues to 
the MW in the local volume.  M101 with an extremely small stellar 
halo \citep{vanDokkum2014,Merritt2016} might be a MW-equivalent in 
terms of its sparse accretion history. 

On the other hand, the MW halo also has a very low stellar metallicity. 
For this work following \cite{Harmsen2017}, we adopt a value of [Fe/H] 
$\sim$ $-$1.7 dex at $\sim$ 30 kpc, which is  the mean metallicity as 
reported by \citet{Sesar2011} and \citet{Xue2015}. Within a heliocentric 
radius of 15 kpc, there is strong evidence of a fairly constant  median 
metallicity \citep{Carollo2007,deJong2010, Xue2015}.  Even within the 
solar neighbourhood,  the metallicity of the kinematically hot halo 
stars is consistent with a flat metallicity gradient of accreted stars 
\citep{Beers2000}. Further out beyond 15 kpc, there seems to be a  
tension between the results of \cite{Sesar2011} who find no metallicity 
gradient and  \cite{deJong2010} and \cite{Xue2015} who find a metallicity 
gradient. Nevertheless, for our purposes,  the total  accreted stellar 
metallicity cannot be larger than $-$1.4 dex. This indicates that the 
MW accreted stellar component was built from a number of  smaller progenitors (low 
$\mathrm{frac_{Dom}}$) and the most massive satellite to contribute to 
the accreted stellar component had a stellar mass much smaller than the SMC 
(less than $\log\,\mathrm{M_{Dom\,*}}\,\sim\,7.5$).  The low metallicity of the 
total accreted stellar component also constrains the time of accretion to  greater than 
$z>1$ (see Figure \ref{fig_MW_diversity}).  

We note that the metallicities do not include 
significant contributions from the Sagittarius stream. Indeed, given 
metallicity and mass estimates for Sagittarius and its stream of 
[Fe/H] $\sim -0.7$ \citep{Hyde2015} and $\mathrm{M_{*}} > 10^8 M_{\odot}$ 
\citep{Niederste2010} respectively, Sagittarius will be 
the dominant progenitor of the Milky Way's stellar halo when it 
disrupts completely. 

\subsubsection*{M31} 

The stellar halo of M31, the closest MW-like mass 
spiral galaxy has been well studied over the last decade.  It would be 
interesting to test if the simple modelling adopted in this work can
explain the detailed measurements of the accreted stellar component of  M31.

M31 has a  fairly large stellar halo compared to the MW. The stellar halo 
mass of M31 outside 27 kpc is estimated to be $\sim 1.1\,\times 10^{10}\,
\mathrm{M_{\odot}}$ \citep{Ibata2014}. Following \cite{Harmsen2017}, 
we assume the total accreted stellar mass of M31 is $\mathrm{M_{Acc,M31}}\sim 
1.5 \pm 0.5 \times 10^{10}\, \mathrm{M_{\odot}}$. \cite{Gilbert2014} estimated 
a metallicity gradient along the minor axis between 10 and 90 kpc as 
$-0.01\,\mathrm{dex\,kpc^{-1}}$.  The metallicity of the stellar halo of M31 at 
at 30 kpc is estimated to be $\mathrm{[Fe/H]}\sim\,-0.65$ \citep{Gilbert2014} 
assuming a 10 Gyrs population. Extrapolating the metallicity gradient, the 
metallicity of the total accreted  stellar component of M31 is estimated 
to be $\mathrm{[Fe/H]}\sim\,-0.3$, which suggests that the stellar mass 
of the most dominant progenitor was $\log\,\mathrm{M_{Dom\,*}}\,\sim\,10.0$.
Moreover, the reasonably steep metallicity gradient along the 
minor axis given M31's accreted stellar metallicity suggests that the dominant progenitor 
was accreted within a lookback time of 4-5 gigayears (see Figure \ref{fig_MW_diversity}).

A prominent feature of the  inner spheroidal halo is the presence of a giant 
stellar stream (GSS) \citep{Ibata2001}.  It would be interesting to speculate
whether the progenitor of the GSS is the same as the  dominant progenitor
we predict through our models. Over the last decade, through a combination 
of TRGB line-of-sight distances \citep{McConnachie2003} and radial velocity 
measurements of the tidal debris of the GSS \citep{Ibata2004,Gilbert2007,Gilbert2009},  
the possible orbit of the  GSS progenitor has been identified, showing that it 
fell almost straight into  M31 from behind on a highly radial orbit.  Moreover, 
the combination of velocity measurements, population studies and dynamical models 
have been able to demonstrate that a number of the tidal features of the inner 
spheroidal halo like the ''West Shelf'' and the ''North-East Shelf'' are caustics 
derived from the leading trail of the progenitor of the GSS \citep{Font2008,
Fardal2006, Fardal2007, Fardal2008, Mori2008,  Fardal2013, Sadoun2014}. 
While the dynamical models differ in various aspects, they generally agree on 
the fact that the progenitor’s initial stellar mass was in the range 
1--5$\times10^{9}\, M_{\odot}$. Detailed population studies of the GSS
indicate that its star formation history is very similar to that of the
inner spheroidal halo \citep{Brown2006}. Moreover, the star formation of 
the progenitor of the GSS shut off around 5 Gyrs \citep{Brown2006,Bernard2015}. 
This is commonly interpreted as the time the GSS progenitor entered the main 
halo of  M31.  The similarities between the predictions of our model
and a possible progenitor of the GSS are encouraging; we postpone a more 
comprehensive analysis to a future publication. 

\subsubsection*{Cen A (NGC 5128)} 
Cen A is the nearest well studied giant elliptical galaxy with  detailed HST 
observations of its stellar halo extending out to  large distances 
(25 $\mathrm{R_{e}}\,\sim\,150\,\mathrm{kpc}$; \citealp{Rejkuba2014}). 
A metal rich component dominates the stellar halo out to the most radially 
extended fields ([Fe/H] $>$ $-$1 dex). It also exhibits a clear metallicity  
gradient ($\Delta\mathrm{[Fe/H]/\Delta\,R\,\simeq\,-0.0054\,dex/kpc}$) out to 
25 $\mathrm{R_{e}}$. 

As is typical of elliptical galaxies, it is difficult to estimate its 
accreted stellar mass, owing to our inability to distinguish the 
\emph{in situ} from the accreted stellar component.  Assuming that
the  stellar populations beyond 60 kpc are predominately accreted 
material, we can  obtain estimates of the mass of the most dominant
progenitor as well as the total accreted stellar mass.

Extrapolating the outer metallicity gradient of Cen A  (at 60 kpc), 
we obtain a total accreted stellar metallicity of $\mathrm{[M/H]}\, 
\sim\, -0.05$ dex, following Fig.\ \ref{fig19} (or $\mathrm{[Fe/H]}\, 
\sim\, -0.31$ dex, taking into consideration an assumed alpha enhancement  
of $[\alpha/\mathrm{Fe}]=0.3$). The uncertainty in the total accreted 
stellar metallicity is around 0.20 dex. Using the metallicity--stellar 
mass relationship \citep{Gallazzi2005}, this suggests that Cen A accreted 
a  massive progenitor with $\log\,\mathrm{M_{Dom\,*}}\,\sim\,10.0 \pm 0.3$. 

Integrating the outer stellar light of Cen A, we find a total stellar 
mass  of $\sim 5 \times 10^9 \, \mathrm{M_{\odot}}$ between 60 and 150 kpc along the major
axis.  Using model of the accreted stellar component of NGC 5128-equivalent 
galaxies in the Illustris simulations (similar to Section \ref{sec:minor}), 
we find that the total accreted stellar mass is 8 times the aperture 
measurement between 60 and 150 kpc, albeit with a larger 0.3 dex scatter.
This suggests that the accreted stellar mass  of  NGC5128 is  
$\log\, \mathrm{M_{acc}} \, \sim \, 10.6 \pm 0.3$. While extrapolating the 
aperture stellar mass provides an upper limit for the accreted stellar
mass, the  mass of the dominant progenitor provides a lower limit. This
is the first time that the accreted stellar mass of an elliptical galaxy
has been constrained to  our knowledge. It is difficult to determine 
constraints on the time of accretion based on the density and metallicity
gradients as we do not have observations of a larger sample of Cen 
A-type  galaxies to compare with.

\subsection{Crafting a way forward}
Although we have measurements of the galaxy metallicity-stellar mass relationship 
over a large range in stellar mass \citep{Gallazzi2005,Kirby2013}, a particular 
lacuna exists around existing measurements for progenitors of MW-like mass 
galaxies ($8.5 <  \log\, M_{*} < 10.5$). More effort needs to be invested in 
constraining the  stellar metallicities  of a volume limited sample of LMC/SMC 
like objects.

The evolution in metallicity-stellar mass relationship could be inferred using 
the accreted stellar metallicities and masses of a bigger volume-limited sample 
of MW-like galaxies. Yet, an independent measure of the evolution of the stellar 
metallicities as a function of stellar mass at high redshift ($z \sim 2$) would 
be very valuable, and would considerably strengthen the robustness of analyses 
of stellar halos. Larger upcoming optical surveys like VANDELS (a deep VIMOS survey of  CANDELS  
UDS and  CDFS fields; \url{http://vandels.inaf.it}) or LEGA-C \citep{vanderWel2016} 
will target star forming galaxies at a range of redshifts up to $z>2.5$ and shed 
greater light on the evolution of the galaxy metallicity-stellar mass relationship.

Although we have used the large number statistics of the Illustris simulations to
our advantage, at some point, we are limited by its resolution. Higher resolution
simulations are needed to confirm these results as well as better calibrate how
we can extrapolate from aperture measurements of the stellar halo to the global 
measurements of the accreted stellar component of galaxies.

We did not consider in this work the information encoded in the velocity distribution of 
the accreted stellar component. Using dissipationless particle-tagging simulations,
\cite{Amorisco2017} suggests that the angular momentum of the most massive 
accretion events is consumed and diluted by dynamical friction, resulting 
in almost non-rotating  contributions with a strong radial bias. On the 
other hand, material deposited by low-mass satellites retain a significant 
 amount of ordered rotations in the outer parts of galaxies. Taking these 
results at face value, this suggests that the velocities of outer halo 
stars may be able to inform us about smaller accretion events; larger 
progenitors may be more affected by the disk potential and dissipation 
\citep[see, e.g.,][]{Gomez2017}.  It will be important to explore kinematic 
signatures of accretion for a wider range of halo masses, and using 
simulations with baryons, incorporating both dissipation and the potential 
from the disk of the main galaxy. 

While the bulk properties of the accreted stellar component are shaped by the most dominant 
progenitor, the presence of substructure in the stellar halo can further inform us 
about the properties of smaller and more recent accretion events \citep[e.g.,][]{Johnston2008}. 
Quantifying faint substructure requires wider area coverage at deep surface brightness limits 
than characterization of the bulk stellar halo, and will benefit from surveys with Hyper 
Suprime-Cam \citep{Miyazaki2012}, LSST \citep[Large Synoptic Survey Telescope][]{arxiv0912.0201L} 
and WFIRST (Wide Field Infrared Survey Telescope) \footnote{\url{https://wfirst.gsfc.nasa.gov/science/sdt_public/WFIRST-AFTA_SDT_Report_150310_Final.pdf}}.  

Constraints on star formation histories may be of particular value in enriching our 
view of galactic accretion histories. While intermediate age AGB stars are bright and may 
be useful near-term probes for intermediate stellar populations for nearby galaxies
\citep{Greggio2014}, most probes of population age will require the advent of larger 
aperture telescopes. The presence of particular old stellar populations (like RR-Lyrae 
or BHB stars) contain further information about older accretion events \citep{Bell2010,Deason2012}. 
Alpha-abundances and their radial trends contain information about star formation 
timescales, but require deep spectroscopy for accurate measurement. 

\section{Conclusions}
\label{sec:concl}
In this paper, we demonstrate using the Illustris simulations the existence of a 
generalized accreted metallicity-stellar mass relationship, extending over three 
orders in magnitude in accreted stellar mass. This relationship is similar to 
the galaxy metallicity-stellar mass relationship, and is offset lower 
by $\sim$0.3 dex.

The accreted metallicity-stellar mass relationship arises because the metallicity
of the stellar halo is driven by the dominant progenitor. More massive satellites 
have a higher metallicity due to the galaxy metallicity-stellar mass relationship. 
Moreover, the dominant progenitor contributes the bulk of the stellar mass to the 
accreted stellar component.
Galaxies with a similar DM halo mass range occupy a steeper locus across the 
accreted metallicity-stellar mass relationship.

We find that the scatter in accreted stellar metallicity at a fixed accreted stellar 
mass encodes information information about the mass of the dominant progenitor. On the
other hand, we find that observable quantities containing spatial information (e.g. 
half-mass radius, density profile, metallicity gradients) of the accreted stellar 
halo encodes information about the time of accretion of the dominant progenitor.

With the goal of using these metrics of galaxy merger history for nearby galaxies, 
we explore possible methods for connecting minor axis resolved stellar population 
observations with model predictions. We can reconstruct the median total accreted 
stellar metallicity of a galaxy to an accuracy of 0.1 dex using a measurement of 
the accreted stellar metallicity along the minor axis of edge-on systems coupled 
with the metallicity gradients at large galacto-centric distances. ``Aperture" 
measurements of the stellar halo mass can be extrapolated to the total accreted 
stellar mass to within an accuracy of 0.15 dex, with further increases in accuracy 
possible by using information encoded in the accreted density profile. We conclude 
that minor axis observations of resolved populations in stellar halos indeed hold 
considerable promise for empirically measuring the mass and timing of the dominant 
accretion event. 

We explore the consequences of  our models on the accretion histories of the MW,
M31 and  NGC 5128. We find the most massive satellite to contribute to 
the accreted stellar component of the MW (excepting Sagittarius) had a stellar mass 
much smaller than the SMC  (less than $\log\,\mathrm{M_{Dom\,*}}\,\sim\,7.5$) and 
was accreted earlier that $z\sim 1$. On the other hand, our models predict that  
M31 accreted a satellite of stellar mass larger than $\log\,\mathrm{M_{Dom\,*}}\,
\sim\,9.7 \pm 0.2$ about 4-5 gigayears ago. We
also constrain the stellar mass of the dominant progenitor and the accreted
stellar mass of NGC 5128 as  $\log\,\mathrm{M_{Dom\,*}}\,\sim\,10 \pm 0.3$ and  
$\log\, \mathrm{M_{acc}} \, \sim \, 10.6 \pm 0.3$ respectively; to our knowledge 
this is the first estimate to have been made of the accreted stellar mass of NGC 5128.

\section*{Acknowledgements}

We thank Alis Deason for providing the properties of her modelled stellar halos in
electronic form. We acknowledge useful discussions with Antonela Monachesi, Kohei
Hattori, Sarah Loebman, Monica Valluri, Ian Roederer and Adam Smercina. We also would 
like to thank the Illustris simulations team, in particular Annalisa Pillepich, Vincent 
Rodrigues-Gomez  and Dylan Nelson.
\nocite{*}

\bibliography{mass_metallicity}
\bibliographystyle{mn2e}


\label{lastpage}

\end{document}